\documentclass[a4paper,12pt,onecolumn]{article}

\usepackage[utf8]{inputenc}
\usepackage[T1]{fontenc}

\usepackage[ngerman]{babel}

\usepackage{FAS}
\usepackage{graphicx} 
\usepackage{csquotes} 
\usepackage{xcolor} 
\usepackage{todonotes} 
\usepackage{subfig} 
\usepackage{ragged2e}
\usepackage{amsthm}
\usepackage{verbatim}
\usepackage{array}
\usepackage{caption}
\usepackage{hyperref}
\usepackage{enumerate}
\usepackage[shortlabels]{enumitem}
\usepackage{microtype}
\usepackage[all]{nowidow}
\usepackage[
			bibstyle=ieee,
			citestyle=numeric-comp,
			backend=biber,
			minnames=1,
			maxcitenames=6,
			maxbibnames=10,
			isbn=false,
			eprint=false,
			url=false,
			natbib=true,
			clearlang=true,
			date=year
			]{biblatex} 
\usepackage{longtable}

\usepackage{ngerman}

\newtheorem*{obersatz}{Obersatz}
\newtheorem*{definition}{Definition}
\newtheorem*{subsumtion}{Subsumtion}
\newtheorem*{ergebnis}{Ergebnis}

\begin{document}
%
\title{Ein Beitrag zur durchgängigen, formalen Verhaltensspezifikation automatisierter Straßenfahrzeuge\footnote{Das Projekt \enquote{Verifikation und Validierung autonomer Fahrzeuge L4/L5} ist Teil der VDA Leitinitiative und Teil der PEGASUS Projekt Familie, vom Bundesministerium für Wirtschaft und Klimaschutz (BMWK, \url{http://www.bmwi.de}) gefördert. Web: \url{www.vvm-projekt.de}}}
%
%
\author{Nayel Fabian Salem%
\footnote{Korrespondierender Autor: salem@ifr.ing.tu-bs.de}
\footnote{Institut für Regelungstechnik, Technische Universität Braunschweig, Braunschweig}
, Veronica Haber%
\footnote{PROSTEP AG, München}
%
%
, Matthias Rauschenbach%
\footnote{Fraunhofer-Institut für Betriebsfestigkeit und Systemzuverlässigkeit LBF, Darmstadt}
,\\
Marcus Nolte%
\footnotemark[3]
, Jan Reich%
\footnote{Fraunhofer-Institut für Experimentelles Software Engineering IESE, Kaiserslautern}
,
Torben Stolte%
\footnotemark[3]
, \\ Robert Graubohm%
\footnotemark[3]
\ und Markus Maurer%
\footnotemark[3]
}
%
%
\date{}

\maketitle \thispagestyle{empty}

\begin{abstract}
    Die Absicherung automatisierter Straßenfahrzeuge (SAE Level 3+) setzt die Spezifikation und Überprüfung des Verhaltens eines Fahrzeugs in seiner Betriebsumgebung voraus. 
    Mithilfe von Szenarien kann der offene Verkehrskontext, in dem ein solches System agiert, strukturiert beschrieben werden. 
    Um Annahmen, welche bei der Verhaltensspezifikation innerhalb von Szenarien getroffen werden, begründen und belegen zu können, ist eine durchgängige Dokumentation dieser Entwurfsentscheidungen erforderlich.
    Mit der Einführung der \textit{semantischen Normverhaltensanalyse} wird eine Methode vorgeschlagen, mithilfe derer Ansprüche an das Verhalten eines automatisierten Fahrzeugs in seiner Betriebsumgebung durchgängig auf ein formales Regelsystem aus semantischen Konzepten für ausgewählte Szenarien abgebildet werden können.
    Der Beitrag der vorgestellten Methode besteht in der Rückverfolgbarkeit dieser formalisierten Konzepte zu den dazugehörigen Ansprüchen an das Verhalten.
    Eine semantische Normverhaltensanalyse wird in dieser Arbeit in zwei ausgewählten Szenarien durchgeführt. Hierfür werden Verhaltensregeln aus einem Auszug der Straßenverkehrsordnung exemplarisch formalisiert.
\end{abstract}


\section{Einleitung}
\label{sec:intro}
Die fortschreitende Automatisierung der Fahraufgabe stellt Beteiligte aus Gesetzgebung, Technik und Ethik gleichermaßen vor neue Herausforderungen. Um die mit dem automatisierten Fahren (SAE Level 3+~\cite{sae_sae_2021}) verbundenen Chancen -- wie beispielsweise die Erhöhung der Verkehrssicherheit -- nutzen zu können, müssen Anforderungen an sicheres und regelkonformes Verhalten automatisierter Straßenfahrzeuge im Straßenverkehr formuliert und geprüft werden.

Unternehmen, welche diese sicherheitskritischen Systeme freigeben, müssen in die Lage versetzt werden, Risiken, welche mit einer Teilnahme am offenen Straßenverkehr assoziiert sind, hinsichtlich ihrer gesellschaftlichen Akzeptanz zu bewerten \cite{wachenfeld_release_2016}. 
Für Entwickler*innen automatisierter Straßenfahrzeuge ergibt sich daraus unter anderem die Aufgabe, zu argumentieren, warum sich das Fahrzeug regelkonform und sicher im Straßenverkehr verhalten wird. 
Um regelkonformes und sicheres Verhalten argumentieren und belegen zu können, wird es, entsprechend des etablierten Sicherheitslebenszyklus der funktionalen Sicherheit nach ISO~26262 \cite{noauthor_road_2018}, notwendig sein, Anforderungen zunächst zu formulieren und anschließend zu überprüfen. Für die Entwicklung automatisierter Straßenfahrzeuge bedeutet dies -- insbesondere für eine Konformität mit der ISO/DIS~21448~\cite{noauthor_road_2021} -- eine Spezifikation des zu implementierenden Verhaltens in Bezug zu seiner Betriebsumgebung.

Eine besondere Herausforderung bei der Verhaltensspezifikation automatisierter Straßenfahrzeuge ergibt sich aus dem Betrieb dieser Systeme im offenen Verkehrskontext.
Dieser offene Kontext, in dem ein automatisiertes Fahrzeug agiert, bedingt eine unbegrenzte Anzahl zu beherrschender Szenarien.
Der szenarienbasierte Ansatz zur Strukturierung der Betriebsumgebung automatisierter Fahrzeuge geht demnach inhärent mit der Unvollständigkeit spezifischer Anforderungen an ihr Verhalten in Bezug auf die betrachteten Szenarien einher \cite{nolte_representing_2018}.
Im Rahmen der Sicherheitsargumentation bedeutet die Unvollständigkeit der Anforderungen, dass 
für eine Bewertung möglicher Restrisiken die getroffenen Annahmen begründet und belegt werden müssen \cite{stellet_formalisation_2019}.

Um einen Beitrag zur Argumentation einer sicheren und konformen Verhaltensspezifikation für einen Szenarienkatalog zu leisten, wird in dieser Arbeit die Durchgängigkeit bei der Erhebung und Dokumentation von Anforderungen an das Verhalten automatisierter Straßenfahrzeuge fokussiert.
Die im Folgenden vorgestellte \textit{semantische Normverhaltensanalyse} unterstützt den Prozess der expliziten Dokumentation von Annahmen und getroffenen Entwurfsentscheidungen bei der Spezifikation von zu implementierendem Verhalten durch die Formalisierung von Verhaltensregeln für betrachtete Szenarien. 
Die Vollständigkeit der resultierenden Verhaltensspezifikation ist in Bezug auf den offenen Kontext weiterhin inhärent nicht gegeben, da die Analyse sich auf einen (idealerweise repräsentativen) Szenarienkatalog bezieht. Der Beitrag der semantischen Normverhaltensanalyse besteht in der Rückverfolgbarkeit und Erklärbarkeit formulierter Verhaltensregeln. 
Eine explizite Dokumentation der Verhaltensregeln sowie damit einhergehender Annahmen und Entwurfsentscheidungen ermöglicht darüber hinaus eine Anpassung auf Basis neuer Erkenntnisse innerhalb des Produktlebenszyklus, wie sie zum Beispiel aus der aktuellen Rechtsprechung oder Unfalluntersuchungen gewonnen werden können. Diese können anschließend explizit in die Weiterentwicklung des automatisierten Fahrzeugs einbezogen werden.

Im folgenden Abschnitt\,\ref{sec:term} wird die im Rahmen dieses Beitrags verwendete Terminologie eingeführt. Anschließend (Abschnitt\,\ref{sec:relwork}) wird der vorgestellte Ansatz zu bestehenden wissenschaftlichen Arbeiten abgegrenzt und in den durch die ISO/DIS~21448 \cite{noauthor_road_2021} und ISO~26262 \cite{noauthor_road_2018} teilweise vorliegenden normativen Rahmen eingeordnet. 
Anschließend wird der methodische Ansatz erläutert (Abschnitt\,\ref{sec:meth}) und anhand eines Fallbeispiels seine Anwendung demonstriert (Abschnitt\,\ref{sec:bsp}). 
An die beispielhafte Demonstration schließt sich eine Evaluation an (Abschnitt\,\ref{sec:eval}), in der Grenzen des Ansatzes diskutiert und geplante zukünftige Arbeiten zur semantischen Normverhaltensanalyse skizziert werden.

\section{Terminologie}
\label{sec:term}
Als terminologische Grundlage für den vorgestellten Ansatz werden die Begriffe \textit{Normverhalten} und \textit{Sollverhalten} definiert als:

\begin{quote}
        {Das \textbf{Normverhalten} ist das aus gesetzlichen, gesellschaftlichen und ethischen Regeln sowie Sicherheitsmechanismen resultierende Verhalten eines Akteurs in einem \textbf{szenarienübergreifenden Kontext}.}
\end{quote}

\begin{quote}
        {Das \textbf{Sollverhalten} ist das sich aus gesetzlichen, gesellschaftlichen und ethischen Regeln sowie Sicherheitsmechanismen abgeleitete, umzusetzende Verhalten eines Akteurs in einem \textbf{szenarienspezifischen Kontext.}}
\end{quote}

Der Begriff \textit{Verhalten} folgt hierbei im Gegensatz zu Censi\,u.\,a.~\cite{censi_liability_2019} der Definition im Rahmen der verhaltensbasierten Robotik \cite{arkin_behavior-based_1998, mataric_behavior-based_2008} und beinhaltet dementsprechend neben dem von außen beobachtbaren Verhalten auch die dafür notwendigen Informationsflüsse als Stimuli der Systemantwort des Akteurs.

\begin{figure}[!h]
	\centering
	\captionsetup{width=.8\linewidth}
	\includegraphics[width=0.9\linewidth]{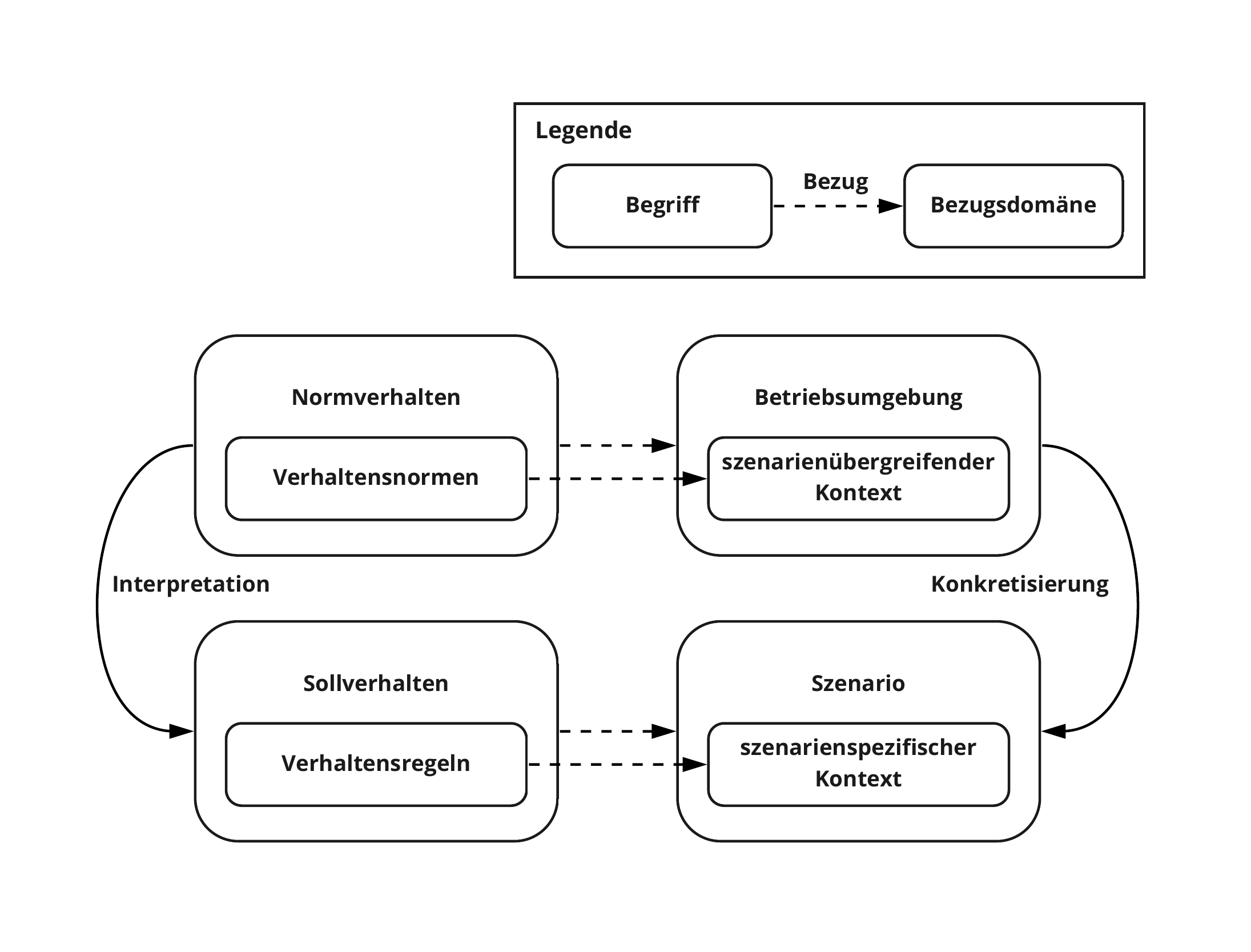}
	\caption{Im Rahmen dieses Beitrags zur semantischen Normverhaltensanalyse werden die Begriffe \textit{Normverhalten} und \textit{Sollverhalten} sowie \textit{Verhaltensnormen} und \textit{Verhaltensregeln} eingeführt und der Zusammenhang mit den jeweiligen Bezugsdomänen beschrieben.}\label{fig:begriffe}
\end{figure}

Im Folgenden werden allgemeine Verhaltensnormen in Summe als Normverhalten und konkrete Verhaltensregeln als Sollverhalten bezeichnet (Abbildung \ref{fig:begriffe}). Die Berücksichtigung gegebenenfalls konfliktärer Verhaltensnormen in einem szenarienübergreifenden Kontext beziehungsweise einer Betriebsumgebung~\cite{walden_systems_2015} (Normverhalten) erfordert bei der Formulierung von Verhaltensregeln in einem gegebenen Szenario~\cite{ulbrich_defining_2015} (Sollverhalten) sowohl eine szenarienspezifische Interpretation als auch eine Auflösung und damit verbundene Abwägung von Zielkonflikten. 

Als Verhaltensnorm könnte beispielsweise formuliert werden, dass durchgezogene Fahrstreifenmarkierungen im Allgemeinen nicht zu überfahren sind. In einem Szenario, in dem der eigene Fahrstreifen belegt und der benachbarte Fahrstreifen frei ist, könnte dagegen die Entwurfsentscheidung getroffen werden, eine Verhaltensregel zu formulieren, welche das Überfahren der durchgezogenen Markierung (z.\,B. aufgrund eines Notstands nach §~16~OWiG) begründet. Die Legitimität der Begründung wird gegebenenfalls \textit{ex post} durch entsprechende Institutionen wie Gerichte und Ethikräte festzustellen sein.
Eine Repräsentation dieser mit der Verhaltensspezifikation verbundenen Annahmen und Abwägungen ist für Entwickler*innen insbesondere notwendig, um eine Argumentationsgrundlage gegenüber firmeninternen und -externen Institutionen zu besitzen, falls unsicheres oder nicht regelkonformes Verhalten eines automatisierten Fahrzeugs beobachtet wird. 

Unter Verwendung der vorgestellten Terminologie kann die semantische Normverhaltensanalyse in mehrerlei Hinsicht einen Beitrag zur Verhaltensspezifikation leisten.

\begin{itemize}
    \item Analysen von Verhaltensnormen eines szenarienübergreifenden Kontextes und dabei getroffene Annahmen werden explizit dokumentiert.
    \item Verhaltensnormen werden über eine szenarienbasierte Betrachtung durchgängig als formale Verhaltensregeln abgebildet.
    \item Der spezifizierte Regelkatalog kann hinsichtlich seiner Konsistenz durch die Anwendung formaler Logik überprüft werden.
    \item Die durchgängige Spezifikation von Verhalten ermöglicht eine Überprüfung der Konformität dokumentierter Verhaltensregeln mit analysierten Verhaltensnormen innerhalb betrachteter Szenarien.
\end{itemize}

\section{Verwandte Arbeiten}
\label{sec:relwork}
Vor der detaillierten Einführung der semantischen Normverhaltensanalyse in Abschnitt~\ref{sec:meth} wird in diesem Abschnitt die konzeptionelle Anschlussfähigkeit zu verwandten Arbeiten skizziert sowie die Erweiterung ausgewählter existierender Ansätze herausgearbeitet.

Die in diesem Beitrag vorgestellte Methode zur durchgängigen, formalen Verhaltensspezifikation ist aus der Sicherheitsbetrachtung automatisierter Straßenfahrzeuge (SAE~Level~3+~\cite{sae_sae_2021}) motiviert. Auch Feth\,u.\,a.~\cite{hoshi_multi-aspect_2018} begründen die Notwendigkeit einer \textit{sicheren nominalen Verhaltensspezifikation} (übersetzt von Salem nach Feth\,u.\,a.~\cite{hoshi_multi-aspect_2018}, \glqq safe nominal behavior specification\grqq) in Bezug auf Grenzen der ISO~26262 \cite{noauthor_road_2018} sowie der zum Veröffentlichungszeitpunkt (von \cite{hoshi_multi-aspect_2018}) noch nicht erschienenen ISO/PAS~21448 \cite{noauthor_road_2019}. Die Adressierung funktionaler Sicherheit und der \textit{Sicherheit der beabsichtigten Funktionalität} (übersetzt von Salem nach ISO/PAS~21448, \glqq safety of the intended functionality\grqq) reicht nach Feth\,u.\,a.~\cite{hoshi_multi-aspect_2018} nicht für einen Sicherheitsnachweis automatisierter Straßenfahrzeuge (SAE Level 3+ \cite{sae_sae_2021}) aus. Die Autoren stellen einen hierarchischen Ansatz zur Entwicklung sicherer automatisierter Fahrzeuge vor, welcher auf der abstraktesten Beschreibungsebene das Fahrzeug in seiner Betriebsumgebung abbildet. 
Der nach der PAS erschienene Normentwurf ISO/DIS~21448~\cite{noauthor_road_2018} reflektiert die Verhaltensspezifikation, deren besondere Relevanz in der Sicherheitsbetrachtung automatisierter Straßenfahrzeuge Feth u.a.~\cite{hoshi_multi-aspect_2018} zuvor postulieren, sowie die Auswirkungen von Defiziten in der Spezifikation ausführlich.

Eine Rolle der Verhaltensspezifikation im Rahmen von Analysen der funktionalen Sicherheit zeigen Graubohm\,u.\,a.~\cite{graubohm_towards_2020} innerhalb der Konzeptphase der Entwicklung eines automatisierten Straßenfahrzeugs. Graubohm\,u.\,a.~nutzen das spezifizierte Verhalten, um systematisch Abweichungen vom Sollverhalten und daraus resultierende Gefährdungen zu identifizieren. Die Autoren beschreiben das Verhalten des Fahrzeugs hierfür zwar mit natürlich-sprachlichen Mitteln, schließen aber eine formale Beschreibung nicht aus.

Für Aspekte der Verhaltenssicherheit nach ISO/DIS~21448 stellen Kramer\,u.\,a.~\cite{kramer_identification_2020} eine schlagwortbasierte Gefährdungsidentifikation vor. Dabei beziehen die Autoren sich auf intendierte Fahrmanöver und davon möglicherweise abweichendes Verhalten. Hierbei nehmen Kramer\,u.\,a.~\cite{kramer_identification_2020} die Verhaltensspezifikation als Bestandteil der analysierten Szenarien als gegeben an.

Die semantische Normverhaltensanalyse trägt mit dem Fokus auf Durchgängigkeit zur Rückverfolgbarkeit der Verhaltensspezifikation, wie sie von Graubohm\,u.\,a.~\cite{graubohm_towards_2020} und Kramer\,u.\,a.~\cite{kramer_identification_2020} genutzt werden, bei. Damit ist der hier vorgestellte Ansatz im Entwicklungsprozess in der frühen Phase der Anforderungsanalyse innerhalb der Konzeptphase automatisierter Straßenfahrzeuge zu verorten. Diese Zuordnung trifft sowohl auf am \mbox{V-Modell} orientierte Entwicklungsprozesse zu als auch auf, wie von Graubohm\,u.\,a.~\cite{graubohm_systematic_2017} skizzierte, iterative Entwicklungsprozesse.


Bezogen auf normative Randbedingungen ist die semantische Normverhaltensanalyse wie folgt einzuordnen: In der ISO/DIS~21448 wird die Erstellung einer \textit{funktionalen Spezifikation} (übersetzt von Salem nach ISO/DIS~21448, \glqq functional specification\grqq) gefordert, welche unter anderem die \textit{Fahrregeln} (übersetzt von Salem nach ISO/DIS~21448, \glqq driving policy\grqq) beinhaltet. Die semantische Normverhaltensanalyse ermöglicht eine explizite Dokumentation von Entwurfsentscheidungen bei der Entwicklung solcher Fahrregeln (bzw. des Sollverhaltens). Auch Analysen der funktionalen Sicherheit nach ISO~26262 wie eine Gefährdungsidentifikation können durch eine Spezifikation des Sollverhaltens unterstützt werden \cite{graubohm_towards_2020}.

Zur Repräsentation spezifizierter Verhaltensregeln wird in dieser Arbeit die Verwendung von Ontologien und Regeln der Prädikatenlogik erster Ordnung \cite{horrocks_swrl_2004} vorgeschlagen. Die Nutzung formaler Sprachen erfüllt den Zweck der automatisierten Überprüfung der Verhaltensspezifikation hinsichtlich ihrer Konsistenz wie von Feth\,u.\,a.~\cite{hoshi_multi-aspect_2018} und Bach\,u.\,a.~\cite{bach_model_2016} gefordert.

Bagschik\,u.\,a.~\cite{bagschik_wissensbasierte_2018} verwenden ebenfalls eine Ontologie und semantische Regeln zur wissensbasierten Szenariengenerierung. Grenzen des Ansatzes zeigen Bagschik\,u.\,a.~\cite{bagschik_wissensbasierte_2018} insbesondere bei der Reduzierung des zu betrachtenden Szenarienraums. Die wissensbasierte Szenariengenerierung ermöglicht durch die Kombination semantisch möglicher Rahmenbedingungen die Generierung einer Anzahl von Szenarien, deren Absicherung nur eingeschränkt möglich ist \cite{bagschik_wissensbasierte_2018}. Der vorgestellte Ansatz zur wissensbasierten Szenariengenerierung bietet keine Lösung für die begründete Auswahl von Szenarien im Rahmen einer szenarienbasierten Absicherung. Mithilfe einer formalen Sollverhaltensbeschreibung kann der von Bagschik\,u.\,a.~\cite{bagschik_wissensbasierte_2018} entwickelte Prozess um eine systematische Beschränkung ergänzt werden, da beispielsweise nicht alle potentiell ausführbaren Fahrmanöver eines automatisierten Fahrzeugs in einem Szenario konform mit bestehenden Verhaltensnormen sein müssen, womit das Szenario nicht im abzusichernden Szenarienkatalog enthalten sein muss.

Eine vergleichbare Ergänzung besteht zur \textit{SCODE for Open Context Analysis (SOCA)} nach Butz\,u.\,a.~\cite{butz_soca_2020}. Die Autoren nutzen Zwicky-Boxen, um den möglichen Entscheidungsraum aufzuspannen und bilden anschließend Äquivalenzklassen, um mögliches Verhalten in einem Szenario zusammenzufassen. Die semantische Normverhaltensanalyse kann in diesem Kontext ebenfalls zur begründeten Reduzierung aufgespannter Handlungsoptionen durch die Identifikation regelkonformer Fahrmanöver dienen.



Im Fokus der hier vorgestellten Arbeit steht die Verhaltensspezifikation auf Ebene funktionaler Szenarien. Um Sollverhalten innerhalb von Szenarien dieser Abstraktionsebene beschreiben und analysieren zu können, wurde das Phänomen-Signal-Modell von Beck\,u.\,a.~\cite{beck_phanomen-signal-modell_2021, beck_phenomenon-signal_2022} vorgestellt. Der zugrundeliegende Formalismus greift auf einen Satz formaler Regeln zurück, um Handlungsoptionen im Rahmen des Sollverhaltens zu schlussfolgern. Mithilfe der semantischen Normverhaltensanalyse kann der Prozess zur Spezifikation dieser Verhaltensregeln unterstützt werden.



Dieser Anspruch steht im Unterschied zu anderen Arbeiten, welche Ansätze zur Formalisierung von beispielsweise Verkehrsregeln vorschlagen \cite{rizaldi_formalising_2015, rizaldi_formalising_2017, nikol_formalisierung_2019, maierhofer_formalization_2020, hannah_proposal_2021}. Die zentrale Motivation dieser Ansätze besteht in der direkten Übersetzung von Verkehrsregeln in formale Beschreibungssprachen. Dabei wird der Aspekt der Durchgängigkeit von Konzepten und Regeln in Bezug auf getroffene Entwurfsentscheidungen vernachlässigt. Im Fall von Rizaldi\,u.\,a.~\cite{rizaldi_formalising_2015} und darauf aufbauenden Arbeiten \cite{rizaldi_formalising_2017, nikol_formalisierung_2019, maierhofer_formalization_2020} bleibt weiterhin unklar, wie sich der Ansatz im Kontext einer szenarienbasierten Absicherung anwenden lässt. Hannah\,u.\,a.~\cite{hannah_proposal_2021} ordnen sich mit einem Fokus auf die Harmonisierbarkeit von Prozessen zur Formalsierung von Verkehrsregeln dem szenarienbasierten Ansatz zu, lassen damit aber eine konkrete Umsetzung offen.


In diesem Abschnitt wurden existierende Ansätze, welche eine eine Verhaltensspezifikation automatisierter Straßenfahrzeuge fordern, nutzen oder bereits Vorschläge zu einer möglichen Umsetzung formulieren, ausgewählt vorgestellt und zur semantischen Normverhaltensanalyse abgegrenzt. Während formale Modelle und Methoden zur Dokumentation des angestrebten Verhaltens automatisierter Systeme insbesondere in den Bereichen der Luftfahrt \cite{torens_formally_2019} und Robotik \cite{colledanchise_behavior_2018} aber auch im Rahmen des automatisierten Fahrens \cite{nagel_innervation_2005, censi_liability_2019} bereits existieren, unterscheiden sie sich insbesondere in ihrem expliziten Bezug (bzw. ihrer Durchgängigkeit) zu den zugrundeliegenden Quellen von Annahmen und Verhaltensanforderungen. 
Das automatisierte Fahren im offenen Verkehrskontext stellt mit der Moderation insbesondere gesellschaftlicher und rechtlicher Aspekte \cite{censi_liability_2019} aber auch im Umgang mit Unsicherheiten \cite{nolte_representing_2018} besondere Anforderungen an die Durchgängigkeit einer Verhaltensspezifikation.
Mit der in dieser Arbeit gezeigten, systematischen Ableitung von Verhaltensregeln aus akzeptierten Quellen von Verhaltensnormen wird ein Beitrag zur Durchgängigkeit der Verhaltensspezifikation geleistet.

\section{Ansatz zur semantischen Normverhaltensanalyse}
\label{sec:meth}
Für die Argumentation regelkonformen und sicheren Verhaltens sind Annahmen und Entwurfsentscheidungen bei der Verhaltensspezifikation explizit zu dokumentieren, um \textit{deduktive Lücken} (nach Stellet\,u.\,a.~\cite{stellet_formalisation_2019}, \glqq deductive gaps\grqq) aufdecken und beheben zu können \cite{stellet_formalisation_2019}. Auf Basis einer Analyse und expliziten Beschreibung von Verhaltensnormen in einer Betriebsumgebung erfolgt im vorgestellten Ansatz eine explizite Abbildung in Form maschinenlesbarer Verhaltensregeln in einem Szenario. Daher ist ein Ziel des Ansatzes, Wissen in Bezug auf das Sollverhalten eines Fahrzeugs im Straßenverkehr auf Basis von Normverhalten semantisch zu formalisieren. Die Überprüfung der Verhaltensregeln kann für einen Satz von Szenarien durchgeführt werden. In Bezug auf die Absicherung eines im Anschluss entwickelten, automatisierten Fahrzeugs innerhalb einer Operational Design Domain (ODD) \cite[S. 15]{noauthor_road_2021} und unter Berücksichtigung des offenen Kontexts hängt die Validität des formulierten Sollverhaltens maßgeblich von der Repräsentativität des Szenarienkatalogs ab. Die im Rahmen des Ansatzes formalisierten Verhaltensnormen beschränken sich zunächst explizit auf den betrachteten Szenarienkatalog. Im Rahmen einer Absicherung wird zu zeigen sein, inwiefern Schlussfolgerungen für die gesamte Betriebsumgebung getroffen werden können.

\begin{figure}[!h]
	\centering
	\captionsetup{width=.8\linewidth}
	\includegraphics[width=0.8\linewidth]{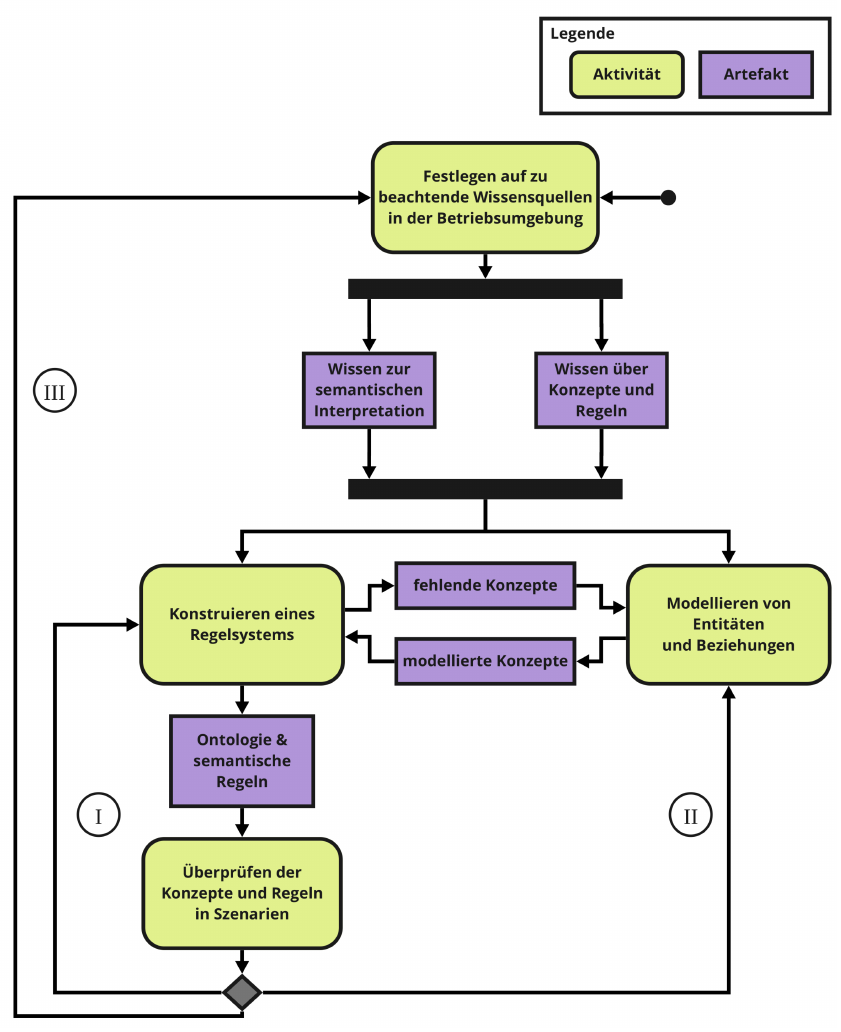}
	\caption{Die Methode der semantischen Normverhaltensanalyse beginnt mit der Auswahl von Wissensquellen, die für die Beschreibung des Normverhaltens in einer Betriebsumgebung genutzt werden. Nach der semantischen Interpretation und Analyse des Normverhaltens ergibt sich ein Regelsystem, welches in ausgewählten Szenarien als modelliertes Sollverhalten überprüft werden kann. Die Iterationsschleifen \RN{1}, \RN{2} und \RN{3} werden durchlaufen, wenn während der Überprüfung des inferierten Sollverhaltens Fehler im Regelsystem, Fehler bei den modellierten Konzepten oder Fehler bei der Auswahl der Wissensquellen festgestellt werden. Fehlerhaftes Verhalten bedeutet im Rahmen des Beitrags nonkormes oder unsicheres Verhalten.}\label{fig:gesamtmethode}
\end{figure}

Abbildung \ref{fig:gesamtmethode} stellt das Vorgehen der semantischen Normverhaltensanalyse dar. Im ersten Schritt werden die einzubeziehenden Wissensquellen wie beispielsweise die Straßenverkehrsordnung (StVO) für die Analyse des Normverhaltens gewählt. Diese Wissensquellen können sowohl Informationen zu Konzepten und Regeln als auch Methoden zum Umgang mit den gewählten Wissensquellen beinhalten. Wissensquellen sind im Kontext der Betriebsumgebung auszuwählen und zu analysieren. Die Betriebsumgebung wird hier verstanden als Summe von Bedingungen, unter denen ein automatisiertes Fahrzeug erwartbar und zulässig betrieben wird.

Um ein Modell des Norm- und Sollverhaltens erstellen zu können, müssen zunächst die Elemente, die für die Beschreibung der Verhaltensregeln notwendig sind, aus den Wissensquellen herausgearbeitet werden.
Während in den meist natürlich-sprach\-lich dokumentierten Quellen teilweise implizites Wissen enthalten ist, kann mit Hilfe expliziter Konzepte erklärbar und rückverfolgbar geschlussfolgert werden. Zum Beispiel verwendet die StVO Konzepte wie Fahrstreifen. Dabei ist ein Fahrstreifen immer Bestandteil einer Fahrbahn. Das Wissen darüber, dass das Befahren eines Fahrstreifens einer Fahrbahn immer auch das Befahren der Fahrbahn bedingt, erscheint zunächst trivial, kann aber bei der formalen Beschreibung eine wesentliche Schlussfolgerung sein.

Für die Formalisierung von Wissen können Ontologien (oder semantische Netze) genutzt werden. In einer Ontologie im informationstechnischen Sinn~\cite{guarino_what_2009} werden Konzepte in Form von Entitäten und Beziehungen modelliert und damit relativ zueinander und nicht absolut definiert. Regeln dienen der Verarbeitung des modellierten Wissens und ermöglichen explizite Inferenzen. Als Konzepte können zum Beispiel ein Fahrzeug, ein Fahrstreifen und eine Straße verstanden werden. Wenn ein Fahrzeug sich auf einem Fahrstreifen befindet und der Fahrstreifen Bestandteil einer Straße ist, kann inferiert werden, dass das Fahrzeug sich auf der Straße befindet. Für diese Schlussfolgerung ist explizit eine Regel zu definieren.
Im Projekt VVMethoden wird an Ontologien zur Repräsentation von Domänenwissen zu Szenarien gearbeitet~\cite{scholtes_6-layer_2021}, die auf Beiträgen im Kontext des Projekts PEGASUS\footnote{https://www.pegasusprojekt.de/de/} aufsetzen~\cite{bagschik_ontology_2018, bagschik_szenarien_2017, menzel_scenarios_2018}.
Die Konzeptualisierung der zu beschreibenden Domäne ist, entsprechend existierender Referenzprozesse zur Entwicklung von Ontologien, der erste Schritt im Anschluss an die Auswahl der Domäne, für die Wissen repräsentiert werden soll~\cite{de_almeida_falbo_sabio_2014, noy_ontology_2001}. In der semantischen Normverhaltensanalyse bilden die Konzeptualisierung von Entitäten und Beziehungen (Schritt 2) sowie die Konstruktion von Regeln (Schritt 3) einen iterativen Prozess, da Kon\-zepte in teilweise implizit repräsentiertem Wissen erst bei der Regelkonstruktion identifiziert werden können.

Wissen über Konzepte und Regeln stammt häufig aus Wissensquellen, die Expertenwissen zur semantischen Interpretation voraussetzen. Daher sind zusätzlich Wissenquellen notwendig, die den methodisch vorgesehenen Umgang mit ersteren Wissensquellen beschreiben.
Diese Methoden werden genutzt, um die Wissensquellen systematisch auf enthaltenes Wissen zu untersuchen und die Konzepte und Regeln zu formalisieren. 
Methodisches sowie domänenbezogenenes Expertenwissen und damit verbundene semantische Interpretationen sind teilweise notwendig, da Entitäten, Relationen und Regeln häufig nicht explizit dokumentiert sind. Ohne Interpretation impliziter zu expliziten Konzepten wäre eine formale Repräsentation semantischer Beziehungen nicht möglich. Das in Abschnitt \ref{sec:bsp} folgende Beispiel erfordert etwa die Repräsentation des Konzepts von \enquote{erkennbar querenden zu Fuß Gehenden}. Die StVO als herangezogene Wissensquelle gibt keinen unmittelbaren Aufschluss darüber, welche Bedingungen an dieses Konzept geknüpft sind. Domänenexperten könnten hier zum Beispiel Gerichtsurteile heranziehen, um eine Interpretation zu stützen.

Eine Anbindung der Wissensmodellierung an existierendes Domänenwissen durch die flexible Wahl der Wissensquellen sowie der Methoden zur semantischen Interpretation ist ein wesentliches Ziel des Ansatzes zur Normverhaltensanalyse. Der hier vorgestellte Ansatz stellt daher die Anforderung an Domänen-Expert*innen, die Wissensquellen bezüglich enthaltener Konzepte und Regeln zu trennen. 
Die Strukturierung von Konzepten und Regeln ermöglicht eine formale Überprüfbarkeit der modellierten Verhaltensregeln hinsichtlich ihrer Konsistenz. Ob und an welche Domänen eine Anschlussfähigkeit gegeben ist, wird die zukünftige Nutzung der semantischen Normverhaltensanalyse zeigen.

Verifiziert wird der erzeugte Regelkatalog in einem letzten Analyseschritt, in dem die Regeln auf alle betrachteten Szenarien angewendet werden. Dieser Schritt kann zum Beispiel mithilfe eines Phänomen-Signal-Modells \cite{beck_phanomen-signal-modell_2021, beck_phenomenon-signal_2022} durchgeführt werden. Mit der semantischen Normverhaltensanalyse wird zunächst nur eine systematische Ableitung von Verhaltensregeln für Szenen \cite{ulbrich_defining_2015} eines funktionalen Szenarios \cite{menzel_scenarios_2018, bagschik_szenarien_2017} beabsichtigt. Das Sollverhalten setzt sich auf der Beschreibungsebene eines funktionalen Szenarios innerhalb eines Phänomen-Signal-Modells anschließend aus den angewendeten Verhaltensregeln zusammen.

Aus der Verifikation ergeben sich im Rahmen der semantischen Normverhaltensanalyse (Abbildung \ref{fig:gesamtmethode}) drei unterschiedliche Arten möglicher Fehlerfälle, die durch drei entsprechende Iterationsschleifen adressiert werden können. Fehlerfälle bezeichnen hierbei eine Abweichung des aus dem spezifierten Regelkatalog resultierenden Verhaltens vom durch die Analyse der Wissensquellen vorgegebenen Verhalten und damit nonkonformes bzw. unsicheres Verhalten. Die erste Iteration (\RN{1}) wird bedingt durch einen Fehlerfall, der sich aus unzureichend definierten Regeln ergibt. Unzureichend bedeutet hier widersprüchlich oder nicht ausreichend detailliert. Ziel dieser Schleife ist die Erzeugung eines konsistenten Regelkatalogs. Der Eintritt in die zweite Iterationsschleife (\RN{2}) wird durch fehlende Konzepte ausgelöst. Diese Schleife ist notwendig, da selbst in einem konsistenten Regelkatalog festgestellt werden kann, dass die Konzeptualisierung für ein Szenario nicht hinreichend vollständig ist. Das Ziel der Schleife ist die Korrektheit des Regelkatalogs in Bezug auf die untersuchten Szenarien. In der dritten und äußersten Schleife (\RN{3}) kann eine in Bezug auf die Szenarien und Wissensquellen hinreichend vollständige Ontologie der Konzepte und ein widerspruchsfreier Regelkatalog vorliegen. Falls dennoch in Frage steht, ob das geschlussfolgerte, regelkonforme Verhalten dem tatsächlich gewünschten Sollverhalten entspricht, kann die Wissensbasis geprüft und es können neue Quellen hinzugezogen werden.

Da Szenarien für die Ableitung von Verhaltensnormen auf Sollverhaltensregeln genutzt werden, ist die Repräsentativität des Szenarienkatalogs maßgeblich verantwortlich für die Validität des formalisierten Sollverhaltens innerhalb einer Betriebsumgebung. Hierbei ist zu beachten, dass in der Betriebsumgebung im Gegensatz zur ODD keine Annahmen zu den Fähigkeiten des automatisierten Fahrzeugs enthalten sind. Das bedeutet, dass ein in allen betrachteten Szenarien valides Sollverhalten nicht auf die Fähigkeiten des Fahrzeugs, dieses Sollverhalten umzusetzen, schließen lässt.

Die Überprüfung der Übertragbarkeit formalisierter Konzepte und Regeln von den analysierten Szenarien auf den offenen Kontext ist nicht Bestandteil des vorgeschlagenen Ansatzes. Grenzen des Ansatzes können daher in Hinblick auf die Repräsentativität des genutzten Szenarienkatalogs und die Validität der verwendeten Wissensquellen identifiziert werden. In Bezug auf die verwendeten Wissensquellen bezieht sich die Validität auf die Akzeptanz der Auswahl aus der Sicht relevanter Stakeholder in der Betriebsumgebung. Zusätzlich kann die fehlerhafte Umsetzung der vorgeschlagenen Methode einen Validitätsverlust bedeuten.

\section{Anwendung des Ansatzes an einem Beispiel}
\label{sec:bsp}
In diesem Abschnitt wird die semantische Normverhaltensanalyse anhand zweier Szenarien beispielhaft demonstriert. Zunächst werden die Szenarien eingeführt und ihre Unterschiede beschrieben. Für die Demonstration des Ansatzes zur semantischen Normverhaltensanalyse werden die zwei ausgewählten Szenarien beispielhaft auf Verhaltensnormen analysiert. Hierbei wird als Beispiel einer Wissensquelle ein Abschnitt der Straßenverkehrsordnung (StVO) \cite{noauthor_strasenverkehrs-ordnung_2013} verwendet. Anschließend wird das Sollverhalten auf Basis des analysierten Normverhaltens formalisiert.

\begin{figure}[!h]
  \centering
  \captionsetup{width=.8\linewidth}
  \subfloat[][]{\includegraphics[width=0.45\linewidth]{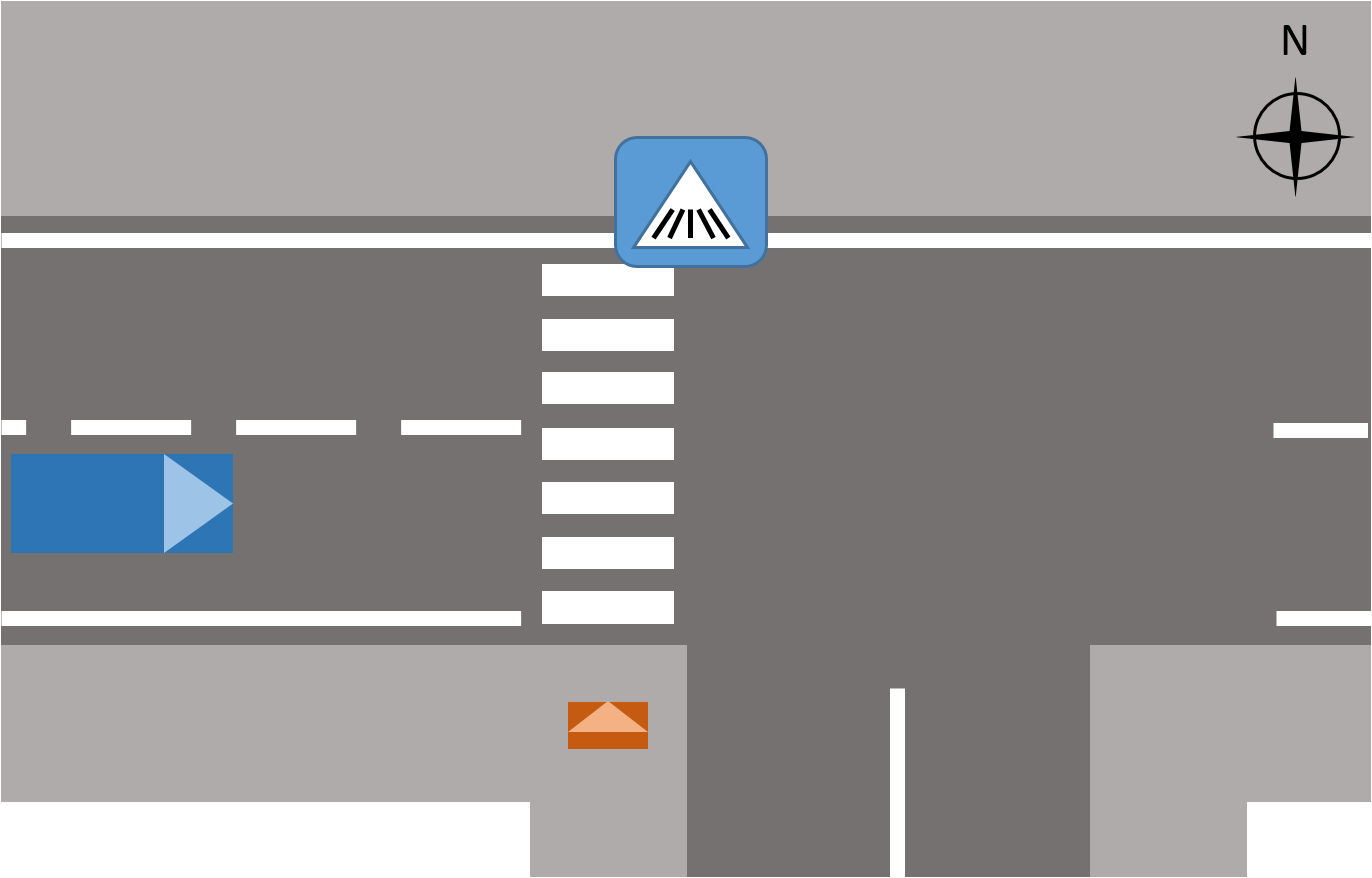}}
  \qquad
  \subfloat[][]{\includegraphics[width=0.45\linewidth]{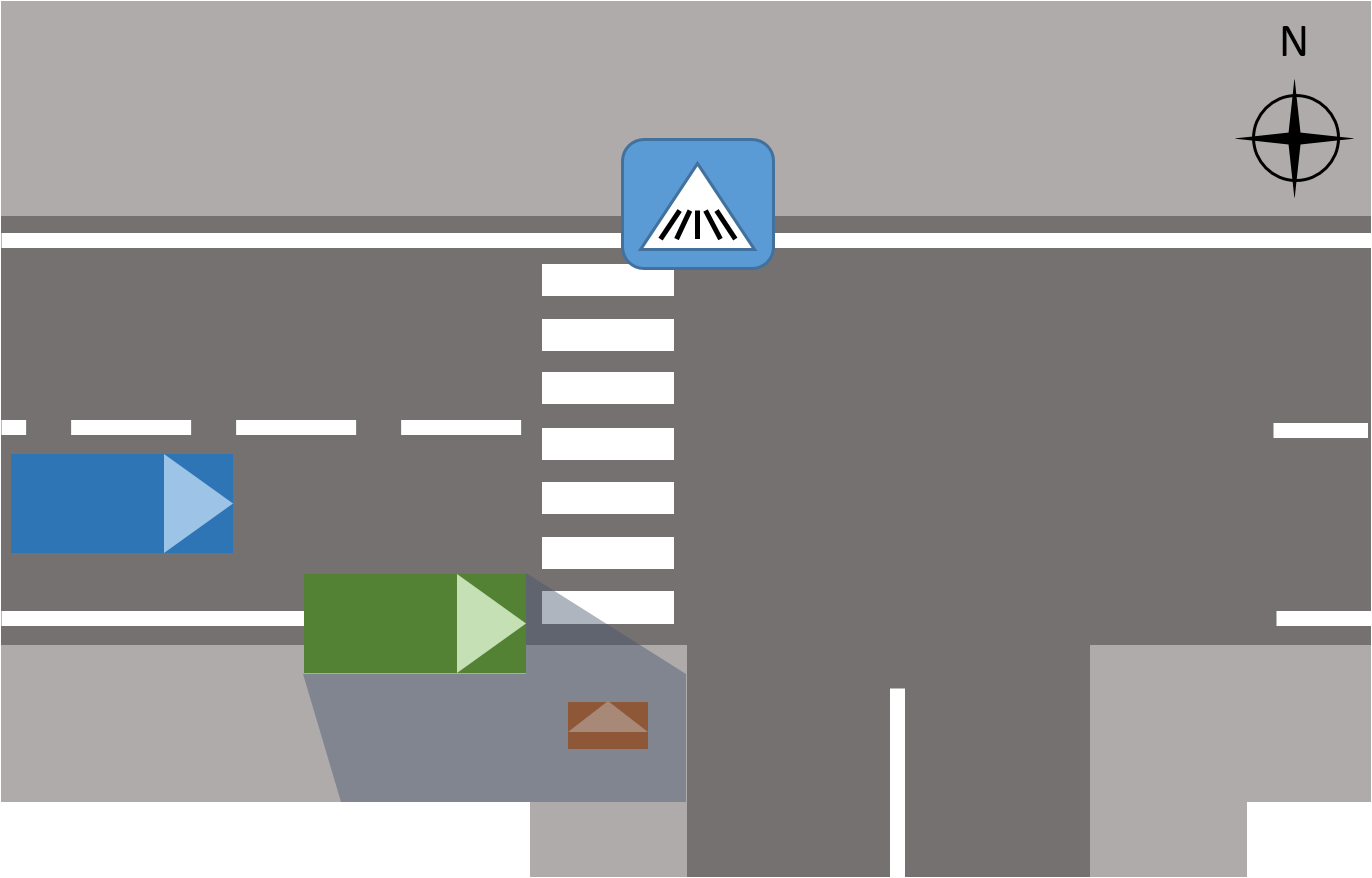}}
  \caption{Die gewählten funktionalen Szenarien beinhalten das automatisierte Ego-Fahrzeug (blau), ein parkendes Fahrzeug (grün) und einen auf den Fußgängerüberweg zulaufenden Fußgänger (orange). Der aus Sicht des Ego-Fahrzeugs verdeckte Bereich wurde zur Illustration ergänzt und ist nicht Teil der Szenenbeschreibung.}\label{fig:fuc2}
\end{figure}

Funktionale Szenarien \cite{menzel_scenarios_2018, bagschik_szenarien_2017} können verwendet werden, um \glqq Betriebsszenarien des Entwicklungsgegenstands auf semantischer Ebene\grqq \cite{bagschik_szenarien_2017} darzustellen.
Abbildung \ref{fig:fuc2} zeigt zwei funktionalen Szenarien, in denen das automatisierte Fahrzeug von Westen in eine T-Kreuzung einfährt und ein Fußgänger von Süden auf einen Fußgängerüberweg der T-Kreuzung zuläuft. In Abbildung \ref{fig:fuc2}a ist der südliche Einlaufbereich des Fußgängerüberwegs vollständig durch das automatisierte Fahrzeug einsehbar. Im Gegensatz dazu behindert ein parkendes Fahrzeug in Abbildung \ref{fig:fuc2}b die Einsicht in den Einlaufbereich bis zum Haltepunkt am Fußgängerüberweg.

Der erste Schritt der semantischen Normverhaltensanalyse (Abbildung \ref{fig:gesamtmethode}) erfordert eine Definition der zu beachtenden Wissensquellen innerhalb der Betriebsumgebung. Da die Definition der Betriebsumgebung selbst nicht Gegenstand dieser Analyse ist, werden ausschließlich notwendige Annahmen für die folgenden Schritte formuliert.

\begin{enumerate}[start=1,label={(\bfseries A\arabic*)}]
    \item Das automatisierte Fahrzeug wird ausschließlich in Deutschland betrieben.
    \item Das automatisierte Fahrzeug wird auf öffentlichen Straßen betrieben.
    \item Das automatisierte Fahrzeug wird im urbanen Raum betrieben. \label{assumption:A3}
\end{enumerate}


\subsection{Exemplarische Rechtsanalyse}

Ausgehend von diesen Annahmen wird als Wissensquelle für die Konzeptualisierung sowie die Formalisierung von Regeln beispielhaft ein Abschnitt der deutschen StVO in der Fassung von 2013 \cite{noauthor_strasenverkehrs-ordnung_2013}
verwendet. Weiterhin wird die Allgemeine Verwaltungsvorschrift
zur Straßen\-verkehrs\-ordnung (VwV-StVO) in der Fassung von 2021 \cite{noauthor_allgemeine_2021}
herangezogen. 

Um Wissen über Konzepte und Regeln, welches in den gewählten Wissensquellen (hier ein Abschnitt der StVO) enthalten ist, semantisch interpretieren zu können, werden in diesem Beispiel Methoden zur Erstellung rechtswissenschaftlicher Gutachten~\cite{puppe_kleine_2019, hildebrand_juristischer_2017} genutzt. Üblicherweise wird bei der Erstellung eines Gutachtens das vorgegebene Szenario (bzw. der \glqq Fall\grqq) mithilfe des gutachterlichen Viererschritts (Obersatz, Definition, Subsumtion und Ergebnis) zerlegt und auf Basis relevanter Rechtsnormen bewertet. Beim verwendeten Gutachtenstil handelt es sich nicht nur um ein Vorgehen, das zur Erstellung von Gutachten angewendet wird, sondern auch um \enquote{[..]eine Art juristisch zu denken}~\cite[S. 1]{hildebrand_juristischer_2017}. Mit dieser Denkweise wird Wissen im Rahmen der semantischen Normverhaltensanalyse interpretiert und systematisch dokumentiert. Das bedeutet, dass hier beispielhaft geleistete Interpretationen der StVO durch den Gutachtenstil und die verwendete Analysemethode maßgeblich die formalisierten Konzepte und Regeln beeinflussen. Gleichzeitig ermöglicht die explizite Interpretation eine durchgängige Argumentation getroffener Annahmen bei der Formalisierung.

Im \textbf{Obersatz} des Gutachtens wird zunächst eine zentrale Fallfrage formuliert, die \enquote{neutral und ohne unwichtige Exkurse}~\cite[S. 2]{hildebrand_juristischer_2017} im Laufe des Gutachtens beantwortet werden soll. Nach Hildebrand~\cite{hildebrand_juristischer_2017} muss das Gutachten dabei vor allem die Eigenschaften der Neutralität und Ökonomie erfüllen. Es ist also eine Wertung durch den Autor zu vermeiden und es sind ausschließlich die für die Beantwortung der Fallfrage notwendigen Informationen im Gutachten zu repräsentieren. 
Die resultierende Anforderung an die Fallfrage im Obersatz ist eine möglichst wertfreie, konkrete Formulierung. Hierbei werden häufig ein oder mehrere zentrale Paragraphen in die Frage hinein formuliert.
Es ist zu beachten, dass Regeln für fahrende Personen auf Regeln für das automatisierte Fahrzeug übertragen werden.
Ausgehend von der Motivation der semantischen Normverhaltensanalyse und dem vorliegenden Szenario könnte der Obersatz für das Anwendungsbeispiel folgendermaßen formuliert werden.
\begin{obersatz}
Welche Pflichten entstehen für die fahrende Person des blauen Fahrzeugs nach \S26 StVO in der in Abbildung \ref{fig:fuc2}a abgebildeten Szene.
\end{obersatz}
Diese Formulierung des Obersatzes ist aufgrund seines konkreten Bezugs nicht ohne Anpassung auf das zweite Szenario in Abbildung \ref{fig:fuc2}b anwendbar. Die Evaluation wird zeigen, dass für jedes zu betrachtende Szenario eine semantische Analyse der Verhaltensnormen notwendig ist, um ein korrektes Sollverhalten abzuleiten.
Üblicherweise wird der Obersatz eines Gutachtens als Entscheidungsfrage formuliert. Im Rahmen dieses Beispiels einer semantischen Normverhaltensanalyse liegt der Fokus auf der Konzeptualisierung und Regelkonstruktion, weshalb die Fragestellung offen gehalten ist.

Im nächsten Schritt des gutachterlichen Viererschritts, der \textbf{Definition}, werden alle relevanten Rechtsnormen, die zur Beantwortung der Fallfrage notwendig sind, herangezogen und so weit analysiert, wie es zur Beantwortung der Fallfrage nötig ist. Bei der Definition werden alle relevanten Tatbestände und Tatbestandsmerkmale gesammelt und erklärt. Tatbestände werden als Grundlage einer Rechtsfolge verstanden. Für einen Tatbestand sind Merkmale definiert, die für dessen Gültigkeit heranzuziehen sind.
Im betrachteten Anwendungsbeispiel könnte ein Teil der Definition lauten:
\begin{definition}
Nach \S26 (1) Satz 1 StVO besteht für die fahrende Person eines Fahrzeugs die Pflicht, \glqq an Fußgängerüberwegen den zu Fuß Gehenden, [...], welche den Fußgängerüberweg erkennbar nutzen wollen, [...], das Überqueren der Fahrbahn zu ermöglichen.\grqq
\\
Laut der VwV-StVO zu \S26 StVO IV. erfolgt die Kennzeichnung eines Fußgängerüberwegs mit der Markierung Zeichen 293. Zusätzlich wird durch Zeichen 350 auf Fußgängerüberwege hingewiesen.
\end{definition}
In der Definition wurde herausgearbeitet, dass \S26 StVO sich unter anderem auf den Tatbestand des Vorhandenseins eines Fußgängerüberwegs bezieht.
Die Tatbestandsmerkmale, anhand derer dieser Tatbestand erkannt werden kann, sind in der VwV-StVO zu \S26 StVO definiert.
Obligatorisch für einen gültigen Fußgängerüberweg ist demnach die Markierung \glqq Zeichen 293\grqq~(der \textit{Zebrastreifen}). Zusätzlich weist das blaue Hinweisschild \glqq Zeichen 350\grqq~fahrende Personen auf das Vorhandensein eines Fußgängerüberwegs hin.
Das Beispiel zeigt, dass die Erkenntnisse aus der Definition genutzt werden können, um Konzepte und Schlussregeln für die Wissensmodellierung abzuleiten.
Nachdem durch den Obersatz eine durch die Analyse zu beantwortende Fragestellung vorgegeben wurde, kann der Definitionsschritt einen Beitrag zur expliziten Interpretation von Gesetzestexten leisten. Im Rahmen der exemplarischen Anwendung wurde die Definition als eine wesentliche Schnittstelle zwischen Expert*innen der Wissensquellen und der Anwendungsdomäne identifiziert.

In der \textbf{Subsumtion} des Gutachtens werden das Szenario und das terminologische Gerüst, welches in der Definition aufgespannt wurde, miteinander verbunden. Die abstrakten Schlüsselbegriffe der Definition werden auf das vorliegende Szenario angewendet, indem verglichen wird, ob die im Szenario vorliegenden Sachverhalte anhand der definierten Tatbestandsmerkmale dem in der Gesetzesnorm vorliegenden Tatbestand zugeordnet werden können \cite[S. 26 ff.]{hildebrand_juristischer_2017}):
\begin{subsumtion}
An der in Abbildung \ref{fig:fuc2}a abgebildeten Kreuzung ist eine Straßenmarkierung vorhanden, die sich als Zeichen 293 der StVO einordnen lässt. Zudem enthält das Szenario am Straßenrand neben der Markierung 293 das Schild 350 als Hinweis auf einen Fußgängerüberweg. Somit liegt im Szenario, das in \ref{fig:fuc2}a abgebildet ist, laut VwV-StVO zu \S26 StVO IV. die Kennzeichnung eines Fußgängerüberwegs vor.
\end{subsumtion}
In der Subsumtion werden allgemeine Beschreibungen aus der Definition auf Sachverhalte im vorliegenden Szenario angewendet (syllogistische Schlussfolgerung). Im Beispiel wird aus dem Vorliegen der Markierung und Beschilderung darauf geschlossen, dass dies dazu dient, einen Fußgängerüberweg zu kennzeichnen. Die Eigenschaft der Subsumtion ist dabei, dass durch die Verbindung der abstrakten Beschreibungsebene mit einem Szenario abschließend geklärt werden kann, ob die vorliegenden Sachverhalte mit den Tatbeständen aus der analysierten Rechtsnorm übereinstimmen. Hieraus können Schlussfolgerungen auf das von den Rechtsnormen vorgeschriebene Verhalten in einer bestimmten Situation abgeleitet werden. Die Subsumtion des Gutachtens beschreibt das Sollverhalten für die konkrete Situation auf Basis der in der Definition erstellten Regeln. 
Nach der Formalisierung der Konzepte und Regeln können die auf Basis des mathematisch logisch beschriebenen Regelkatalogs getroffenen Schlussfolgerungen mit den Schlussfolgerungen der Subsumtion verglichen werden. Falls das inferierte Verhalten nicht mit dem rechtlich geforderten übereinstimmt, sind die formalisierten Regeln anzupassen.

Der letzte Schritt bei der Erstellung eines Gutachtens ist das Ausformulieren eines \textbf{Ergebnisses}. Das Ergebnis beantwortet dabei die im Obersatz formulierte Frage. Hierbei wird auf alle Schlussfolgerungen, die in der Subsumtion getroffen wurden, Bezug genommen und ein Gesamtergebnis formuliert:
\begin{ergebnis}
Die fahrende Person des blauen Fahrzeugs aus der in Abbildung \ref{fig:fuc2}a beschriebenen Situation ist nach \S 26 StVO dazu verpflichtet, dem zu Fuß Gehenden das Überqueren der Fahrbahn zu ermöglichen. Somit darf sie nur mit mäßiger Geschwindigkeit heranfahren und muss, wenn nötig, warten.
\end{ergebnis}
Um ein Ergebnis zu formulieren, welches die Frage des Obersatzes abschließend klärt, sind im gewählten Szenario weitere Analysen durchzuführen. 
Im Rahmen des Anwendungsbeispiels wurden hierfür Annahmen bei der Definition und Subsumtion getroffen.
Eine Annahme (basierend auf Annahme \hyperref[assumption:A3]{A3}) innerhalb des Gutachtens war die Markierung des Fußgängerüberwegs innerhalb einer Ortschaft. Diese Annahme wäre im Rahmen einer umfangreicheren Definition zu untersuchen, da sie explizit für die Gültigkeit eines Fußgängerüberwegs gefordert ist.
Weiterhin wurde angenommen, dass der zu Fuß Gehende innerhalb des Szenarios erkennbar den Überweg nutzen möchte. Auch diese Annahme ist im Rahmen eines aussagekräftigen Gutachtens zu untersuchen.
Aus Sicht der semantischen Normverhaltensanalyse ist der Ergebnissatz die Zusammenfassung des Sollverhaltens in Bezug auf das vorliegende Szenario und daher dazu geeignet, als eine Indikation für die Richtigkeit des inferierten Verhaltens zu dienen.

Die beispielhaft gezeigte Anwendung rechtswissenschaftlicher Methoden soll die mögliche Schnittstelle zwischen der Analyse von Gesetzestexten und der technischen Implementierung eines mathematisch logischen Regelwerks im Rahmen der semantischen Normverhaltensanalyse zeigen. Die Übertragbarkeit der umgesetzten Analyse auf andere Wissensquellen und insbesondere andere Rechtskontexte wird im Rahmen dieser Arbeit explizit nicht angenommen. Da die semantische Interpretation von Konzepten und Regeln den Expert*innen der Rechtsdomäne obliegt, wird hier lediglich ein methodischer Vorschlag zur durchgängigen Anbindung an eine formalen Wissensrepräsentation vorgestellt.

\subsection{Formalisierung der Konzepte und Regeln}

Für eine Durchgängigkeit an der Schnittstelle zwischen natürlich-sprachlicher und formaler Repräsentation des Sollverhaltens ist die Wahl des Abstraktionsgrads wesentlich. Im folgenden Teil des Anwendungsbeispiels wird Sollverhalten auf der Ebene funktionaler Szenarien~\cite{menzel_scenarios_2018, bagschik_szenarien_2017} beispielhaft in eine formale Repräsentation übersetzt.

Eine Entscheidung bei der beispielhaften Umsetzung der semantischen Normverhaltensanalyse wird bei der Wahl der genutzten formalen Repräsentations\-sprache getroffen. Die Wahl der Sprache zur Repräsentation des Wissens kann durch unterschiedliche sprachliche Mittel Limitationen bei der Beschreibung analysierter Konzepte und Regeln hervorrufen. In diesem Fall müsste die Repräsentationssprache insofern angepasst werden, dass Wissen aus den Wissenquellen im Sinne der Autor*innen der Wissensquelle abbildbar ist. Im Anwendungsbeispiel werden Konzepte und ihre Beziehungen in der Web Ontology Language (OWL) beschrieben. Diese Ausprägung der Beschreibungslogik eignet sich, um terminologisches Expertenwissen konsistent und maschinenlesbar zu repräsentieren~\cite{mcguinness_owl_2004}. Außerdem wurde das Tool Protégé\footnote{https://protege.stanford.edu/} verwendet, um die Konzepte in der Ontologie zu editieren. Die Nutzung von Protégé kann vor allem durch seine graphische Oberfläche zur Rückverfolgbarkeit zwischen maschinenlesbarem und durch Menschen nachvollziehbarem Wissen beitragen~\cite{gennari_evolution_2003}.

Im Gegensatz zu Rizaldi~u.\,a.~\cite{rizaldi_formalising_2015} wird die begrenzte Expressivität der Logik erster Ordnung nicht als Hindernis bei der Repräsentation des Wissens, sondern als Vorteil für dessen Nachvollziehbarkeit bewertet. Nicht formale Wissensquellen verwenden vorrangig semantisch reiche Konzepte (z.\,B. ein Fußgängerüberweg), die hier mit Hilfe von Beschreibungslogik formalisiert werden. Damit steht die Repräsentation physikalischer Zustandsgrößenverläufe bei der Ableitung des Sollverhaltens während der Konzeptphase nicht im Vordergrund.

\begin{figure}[!h]
	\centering
	\captionsetup{width=.8\linewidth}
	\includegraphics[width=0.5\linewidth]{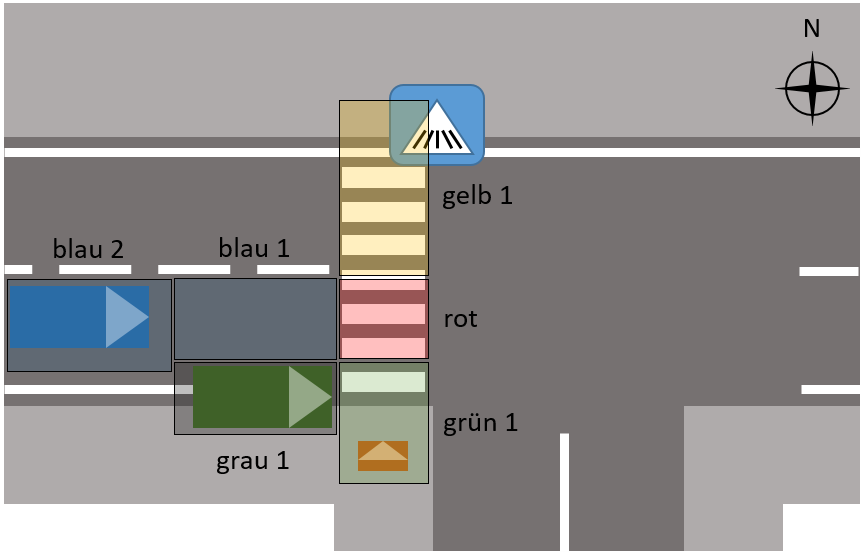}
	\caption{In einem parallel entwickelten Verfahren zur Zonierung einer Szene~\cite{lalitsch-schneider_personliche_2021} werden Zonen auf Basis von Fahrmanövern~\cite{jatzkowski_zum_2021} des Ego-Fahrzeugs und damit verbundenen Fahraufgaben hergeleitet. Zone \enquote{blau 2} und \enquote{blau 1} sind Zonen, in denen das Ego-Fahrzeug zunächst die Geschwindigkeit anpasst und anschließend anhält. In Zone \enquote{rot} kann das Ego-Fahrzeug anfahren und dabei Bezug auf die Zonen \enquote{gelb 1} und \enquote{grün 1} nehmen. Die Zone \enquote{grau 1} ist im Szenario, welches eine Verdeckung beinhaltet, zusätzlich zu berücksichtigen.}
	\label{fig:zonen}
\end{figure}

Eine Eigenschaft der Beschreibungslogik in Bezug auf die Formalisierung von Norm- und Sollverhalten ist, dass sie keine expliziten sprachlichen Mittel zur Repräsentation von Zeit bereitstellt. Diese Eigenschaft wurde adressiert, indem die Konzepte und Regeln bezüglich einer Szene~\cite{ulbrich_defining_2015} formuliert wurden. Da eine detaillierte Szenenrepräsentation im Zusammenhang mit allgemein formulierten Wissensquellen wie der StVO zu impliziten Annahmen führen kann, wurde im Anwendungsbeispiel eine Abstraktion der Szene in Form von Zonen gewählt (Abbildung \ref{fig:zonen}). Anforderungen an die Szenenrepräsentation ergeben sich aus den gewählten Wissensquellen und der semantischen Kompatibilität zwischen abgeleiteten Regeln und formalisierten Konzepten in der Szenenrepräsentation. Ein Beispiel für ein systematisches Vorgehen zur Zonierung einer Szene zeigen Butz~u.\,a.~mit der \textit{SCODE for Open Context Analysis (SOCA)}~\cite{butz_soca_2020}. Alternative Ansätze zur Zonierung von Szenen sind Gegenstand aktueller Forschungsarbeit. Die in diesem Anwendungsbeispiel gewählten Zonen stellen einen Arbeitsstand im Projekt VVMethoden dar \cite{lalitsch-schneider_personliche_2021} und basieren im Wesentlichen auf der Durchführung von Fahrmanövern~\cite{jatzkowski_zum_2021}. Für die semantische Normverhaltensanalyse wird eine Szenenrepräsentation vorausgesetzt, die es ermöglicht, Konzepte und Regeln zu modellieren. Die Zonierung und semantische Normverhaltensanalyse bilden hierbei einen iterativen Prozess, der in diesem Beitrag nicht ausgeführt wird, sondern dessen Ergebnis als gegeben betrachtet wird.

\begin{figure}[!h]
	\centering
	\captionsetup{width=.8\linewidth}
	\includegraphics[width=0.98\linewidth]{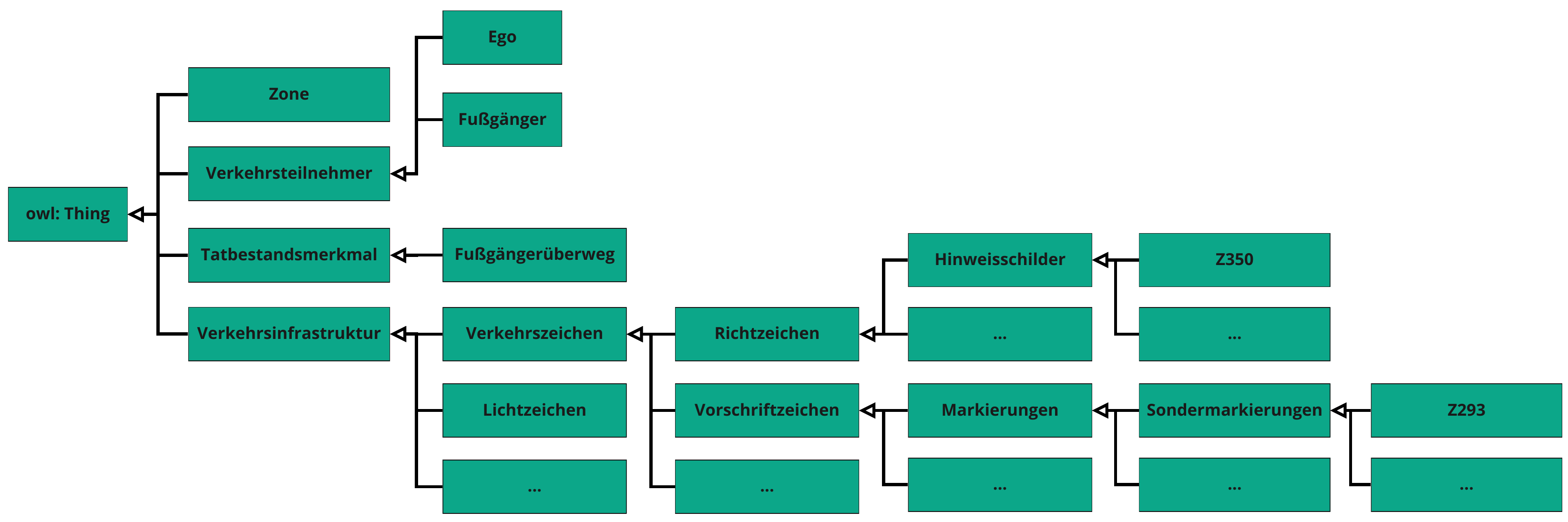}
	\caption{In der TBox (\textit{terminological box}) der Ontologie wurde zunächst eine taxonomische Klassenstruktur aus der StVO abgeleitet und zusätzlich um Klassen erweitert, die für die Szenenrepräsentation benötigt werden.}\label{fig:tbox}
\end{figure}

Bei der ersten Konzeptualisierung der StVO ergibt sich ein in Abbildung \ref{fig:tbox} gezeigter taxonomischer Ausschnitt der Klassenstruktur (oder TBox, \textit{terminological box}). Während die abstraktesten vier Klassen expertenbasiert erzeugt werden, ergibt sich die Dekomposition der Klassenstruktur für die Verkehrsinfrastruktur aus der StVO. Das Konzept des Fußgängerüberwegs wird hier als Tatbestand interpretiert, da an ihn die Rechtsfolge im Sinne des analysierten Abschnitts der StVO geknüpft ist. Zusätzlich kann die Klassenstruktur zum Beispiel durch eine Konzeptualisierung der Betriebsumgebung in Form eines Ebenen-Modells vorgegeben sein~\cite{schuldt_effiziente_2013, bagschik_ontology_2018, scholtes_6-layer_2021}.

\begin{figure}[!h]
	\centering
	\captionsetup{width=.8\linewidth}
	\includegraphics[width=0.75\linewidth]{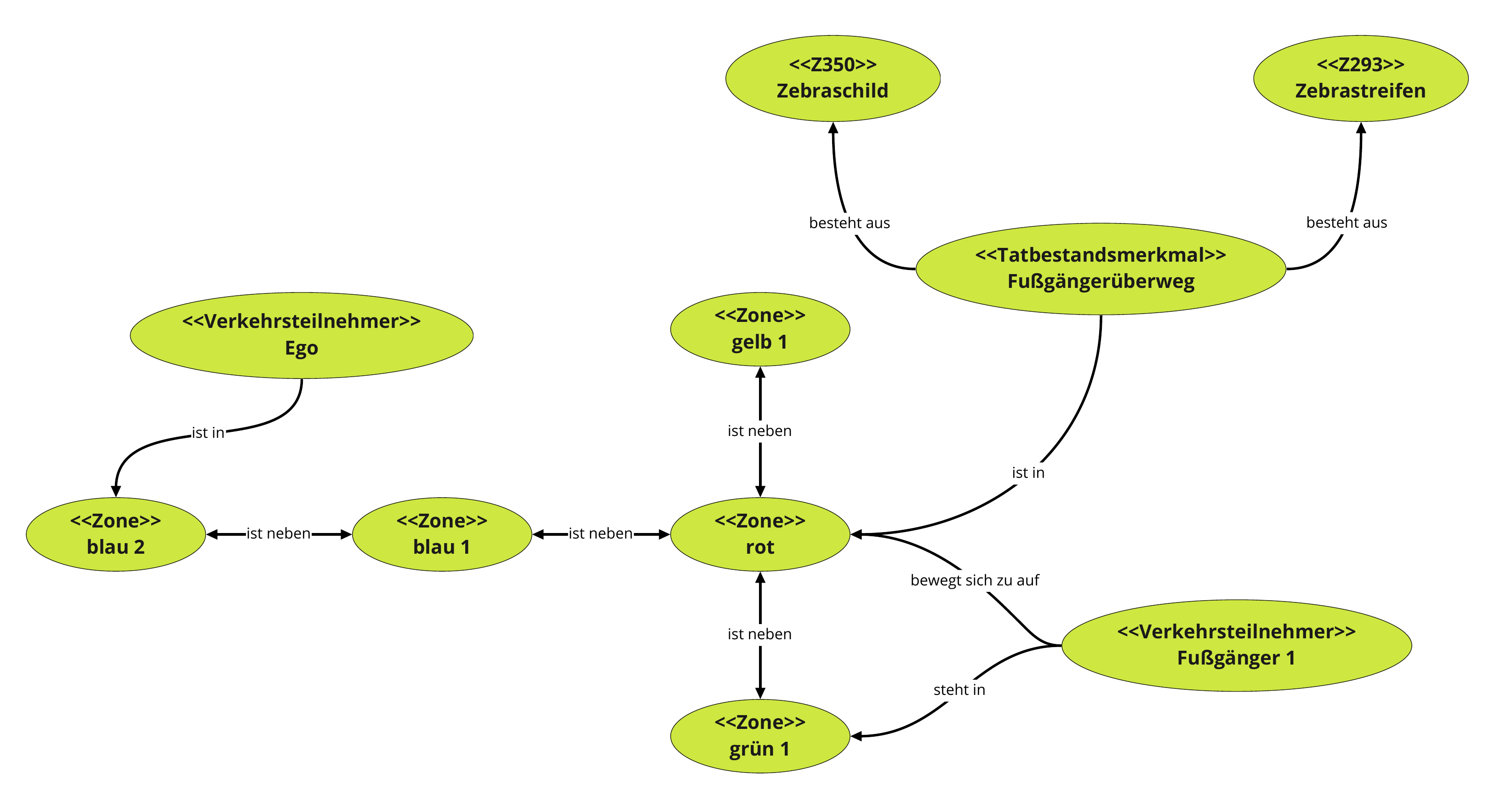}
	\caption{In der ABox (\textit{assertional box}) der Ontologie sind Instanzen der Klassen aus der TBox modelliert, die das erste betrachtete Szenario repräsentieren sollen.}\label{fig:abox}
\end{figure}

Abbildung \ref{fig:abox} zeigt für eine Szene des ersten Szenarios (vgl. Abbildung \ref{fig:fuc2}a) die modellierten Instanzen (ABox, \textit{assertional box}) der Klassenstruktur der Ontologie. In der Szene befindet sich das Ego-Fahrzeug im westlichen Straßenast (Zone \enquote{blau 1}) und fährt in Richtung Osten auf die Kreuzung zu. Der Fußgängerüberweg befindet sich in Zone \enquote{rot} und im Pfad des Ego-Fahrzeugs. Gemäß der StVO besteht ein Fußgängerüberweg aus einem entsprechenden Hinweisschild (Zeichen 350) und einer Markierung (Zeichen 293). Außerdem befindet sich ein Fußgänger im Einlaufbereich des Fußgängerüberwegs (Zone \enquote{grün 1}) und bewegt sich auf ihn zu.

Eine Bottom-up-Vorgehensweise bei der ersten Iteration der Konzeptualisierung von Klassen und Beziehung \textit{(Welche relevanten Konzepte sind in der Szene vorhanden?)} hat sich bei der Entwicklung der Methode bewährt, da auf diese Weise das szenarienbezogene Expertenwissen explizit eingebunden werden kann. Dem gegenüber steht ein Top-down-Vorgehen \textit{(Welche relevanten Konzepte müssen in der Szene repräsentiert sein?)}, welches einer systematischen Abdeckung der zu formalisierenden Konzepte dient. Die Relevanz der gewählten Wissensquellen zeigt sich in diesem Anwendungsbeispiel primär bei der Top-down Konzeptualisierung, da hier die StVO genutzt wurde, um systematisch eine beispielhafte Taxonomie der Verkehrsinfrastrukturelemente abzuleiten (Abbildung \ref{fig:tbox}). In Summe kann mithilfe beider Vorgehensweisen eine Ontologie der Domäne abgeleitet werden, auf die sich der zu formulierende Regelkatalog bezieht.

\begin{figure}[!h]
	\centering
	\captionsetup{width=.8\linewidth}
	\includegraphics[width=0.98\linewidth]{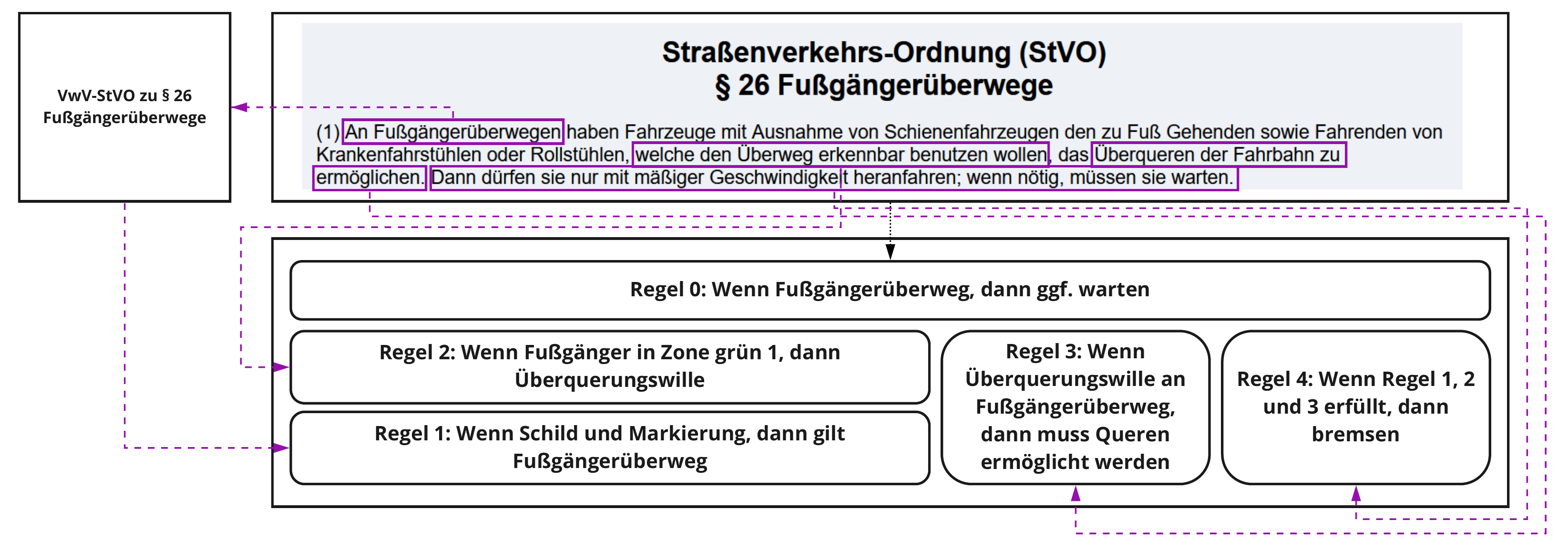}
	\caption{Im Anwendungsbeispiel wird ein Ausschnitt des §~26 StVO mithilfe rechtswissenschaftlicher Methoden analysiert und auf enthaltene Konzepte und Regeln untersucht. Die identifizierten Regeln werden strukturiert und rückverfolgbar mit Ausschnitten des zugrundeliegenden Gesetzestextes verbunden.}\label{fig:regelableitung}
\end{figure}

Anschließend folgt der Schritt der Konstruktion eines Regelkatalogs, der im Ergebnis das Sollverhalten basierend auf den gewählten Wissensquellen in den betrachteten Szenarien formalisiert beschreiben soll. Der Prozess der Konstruktion des Regelkatalogs besteht grundsätzlich aus dem Schritt der Regelableitung und der Regelstrukturierung und ggf. -verfeinerung. Ein Beispiel der rückverfolgbaren Regelkonstruktion basierend auf §~26 (1) StVO ist in Abbildung \ref{fig:regelableitung} dargestellt. In diesem Beispiel wird §~26 (1) StVO auf seine enthaltenen Regeln analysiert. Die abgeleiteten Regeln können zunächst in natürlicher Sprache festgehalten werden.
Bei der Formalisierung der dekomponierten Regeln wird die Notwendigkeit des iterativen Charakters der Methode deutlich. Da Regeln sich teilweise auf implizite Konzepte beziehen, ist es erst bei der Formalisierung der Regel möglich, diese impliziten Konzepte explizit in der Ontologie zu formalisieren. Im gezeigten Beispiel ist das Konzept \enquote{erkennbar queren wollen} nicht explizit beschrieben. Für die Anwendbarkeit der Regel sind dieses Konzept und die damit verbundenen Annahmen zentral, weshalb es für eine semantische Formalisierung explizit beschrieben werden muss.

In dem betrachteten Anwendungsfall wurde §~26 (1) StVO in vier Regeln dekomponiert und formalisiert. Eine Hierarchisierung der Regeln wird in diesem Beispiel nicht gezeigt, da sich aus der StVO keine aufzulösenden Konflikte ergeben. Als \textit{Regel 0} wird hier der Stellvertreter des zu formalisierenden Paragraphen bzw. Absatzes verstanden. Zwischen den Beschreibungsebenen der nicht-formalisierten, natürlich-sprachlichen Wissensquellen und den formalisierten, maschinen-lesbaren Regeln kann der Übersetzungs\-pro\-zess durch im folgenden formulierte semi-formale Regeln unterstützt werden. 

\textit{Regel 0} fasst den herangezogenen Absatz des §~26 StVO zusammen:
\begin{itemize}
    \item \textbf{Wenn} Fußgängerüberweg \textbf{dann} ggf. warten
\end{itemize}

Die Gültigkeit eines Fußgängerüberwegs wird in der VwV-StVO konkretisiert. Das bedeutet, dass \textit{Regel 1} ebenfalls als eine Konkretisierung einer Regelkomponente von \textit{Regel 0} entspricht. In diesem Beispiel wurden die Bedingungen für die Gültigkeit auf das Vorhandensein von Zeichen 293 (Bodenmarkierung Fußgängerüberweg) und Zeichen 350 (Hinweisschild Fußgängerüberweg) reduziert. Dieser Zusammenhang wird in \textit{Regel 1} zusammengefasst:

\begin{itemize}
    \item \textbf{Wenn} Zeichen 293 \textit{und} Zeichen 350 \textbf{dann} gilt Fußgängerüberweg
\end{itemize}

Zudem wird in §~26 (1) StVO beschrieben, dass \enquote{den zu Fuß Gehenden sowie Fahrenden von Krankenfahrstühlen oder Rollstühlen, welche den Überweg erkennbar nutzen wollen} das Überqueren zu ermöglichen ist. In \textit{Regel 2} werden diese Bedingungen vereinfacht formuliert. Die hier getroffene Annahme, dass im betrachteten Szenario die Präsenz einer Person in der Einlaufzone des Fußgängerüberwegs den Willen zum Überqueren ausdrückt, kann durch das Einbeziehen weiterer Wissensquellen wie Gerichtsurteilen konkretisiert bzw. auch widerlegt werden.

\begin{itemize}
    \item \textbf{Wenn} Person in Zone grün 1 (Einlaufzone) \textbf{dann} Person will Überweg erkennbar benutzen
\end{itemize}

Da eine Aktion des Fahrzeugs nur notwendig ist, wenn der Fußgängerüberweg sich im Fahrweg befindet, wird diese Bedingung explizit in \textit{Regel 3} aufgenommen.

\begin{itemize}
    \item \textbf{Wenn} Fußgängerüberweg in Fahrweg \textit{und} Person will erkennbar benutzen \textbf{dann} ist Überqueren zu ermöglichen
\end{itemize}

Um die erwartete Aktion auszudrücken, die mit den vorher beschriebenen Regeln verbunden ist, wird in \textit{Regel 4} die Erfüllung der Bedingungen mit dem Halten in Zone blau 1 verknüpft.

\begin{itemize}
    \item \textbf{Wenn} Regel 1 \textit{und} Regel 2 \textit{und} Regel 3 \textbf{dann} Halten in Zone blau 1 (Haltezone vor Fußgänger\-überweg)
\end{itemize}

Die hier vorgestellte Ableitung von Regeln basiert auf dem Wissen, welches durch das beispielhafte rechtswissenschaftliche Gutachten gewonnen wurde. Im Rahmen der Erstellung des Gutachtens wurde iterativ das Vorhandensein von Konzepten und Regeln geprüft und so die Schnittstelle zwischen der Wissensquelle und der formalen Repäsentation hergestellt. Im Anwendungsbeispiel wird als Notations\-sprache formaler Regeln die \textit{Semantic Web Rule Language (SWRL)}~\cite{horrocks_swrl_2004} verwendet. SWRL ist eine Sprache zur Realisierung von \textit{Description Logic Programs} als Schnittmenge aus Beschreibungslogik und logischer Programmierung in Form von Horn-Formeln~\cite{grosof_description_2003}. 

\begin{table}[h!]
    \centering
    \captionsetup{width=.8\linewidth}
    \caption{Formalisierung semantischer Konzepte in SWRL-Regeln}
    \begin{tabular}{|p{5cm}|>{\raggedright\arraybackslash}p{9cm}|}
         \hline
         \textbf{Natürlich-sprachliche Regel} & \textbf{SWRL-Regel} \\ \hline
         \textbf{Wenn} Zeichen 293 \textit{und} Zeichen 350 \textbf{dann} gilt Fußgängerüberweg &
         \verb+Fussgaengerueberweg(?fuueb)+ $\land$ 
         \verb+Z293_Zebrastreifen(?streif)+ $\land$  \verb+Z350_Fussgaengerueberweg(?schild)+ $\land$ 
         \verb+sachverhalt_gilt(?streif, true)+ $\land$ 
         \verb+sachverhalt_gilt(?schild, true)+ $\land$  \verb+besteht_aus(?fuueb, ?streif)+ $\land$  \verb+besteht_aus(?fuueb, ?schild)+ $->$ \verb+sachverhalt_gilt(?fuueb, true)+ \\ 
         \hline
         \textbf{Wenn} Person in Zone Grün 1 (Einlaufzone) \textbf{dann} Person will Überweg erkennbar benutzen &
         \verb+Zone(?zoneGruen1)+ $\land$ 
         \verb+Fussgaenger(?f2)+ $\land$ 
         \verb+sachverhalt_gilt(?zoneGruen1, true)+ $\land$ 
         \verb+sachverhalt_gilt(?f2, true)+ $\land$ \verb+ist_in(?f2, ?zoneGruen1)+ $->$ \verb+steht_in(?f2, ?zoneGruen1)+\\ \hline
         \textbf{Wenn} Fußgängerüberweg in Fahrweg \textit{und} Person will erkennbar benutzen \textbf{dann} Überqueren zu ermöglichen & 
         \verb+Ego(?ego)+ $\land$ 
         \verb+Fussgaenger(?f2)+ $\land$ 
         \verb+Fussgaengerueberweg(?fuueb)+ $\land$ 
         \verb+Zone(?zoneRot)+ $\land$ 
         \verb+Zone(?zoneGruen1)+ $\land$ 
         \verb+ist_in(?fuueb, ?zoneRot)+ $\land$ 
         \verb+ist_relevant_fuer(?fuueb, ?ego)+ $\land$ 
         \verb+steht_in(?f2, ?zoneGruen1)+ $\land$ 
         \verb+ist_neben(?zoneGruen1, ?zoneRot)+ 
         $->$ 
         \verb+will_evtl_passieren(?f2, ?zoneRot)+\\ \hline
         \textbf{Wenn} Regel 1 \textit{und} Regel 2 \textit{und} Regel 3 \textbf{dann} Halten in Zone Blau 1 (Haltezone vor Fußgängerüberweg) & 
         \verb+Fussgaenger(?f2)+ $\land$ 
         \verb+Fussgaengerueberweg(?fuueb)+ $\land$ 
         \verb+Ego(?ego)+ $\land$ 
         \verb+Zone(?zoneBlau1)+ $\land$ 
         \verb+Zone(?zoneRot)+ $\land$ 
         \verb+will_evtl_passieren(?f2, ?zoneRot)+ $\land$ 
         \verb+ist_Pfadzone(?zoneBlau1, true)+ $\land$ 
         \verb+ist_relevant_fuer(?fuueb, ?ego)+ $\land$ 
         \verb+ist_in(?fuueb, ?zoneRot)+ $\land$ 
         \verb+ist_neben(?zoneBlau1, ?zoneRot)+
         $->$
         \verb+anhalten_in(?ego, ?zoneBlau1)+\\ \hline
    \end{tabular}
    
    \label{tab:swrlregeln}
\end{table}

Die bisher natürlich-sprachlich formulierten Regeln werden in diesem Schritt der Regelkonstruktion in Form von SWRL-Regeln formalisiert. Tabelle \ref{tab:swrlregeln} zeigt die implementierten SWRL-Regeln auf Basis des analysierten Gesetzestexts. Bei der Formulierung der SWRL-Regeln wird hier unterschieden zwischen dem Abfragen der Existenz einer Instanz (z.\,B. \verb+?streif+) einer Klasse (z.\,B. \verb+Z293_Zebrastreifen()+) in der Szene und der Feststellung eines Sachverhalts (z.\,B. \verb+sachverhalt_gilt(?streif, true)+). Im gezeigten Beispiel wurden keine Regeln für die Feststellung der Sachverhalte von zum Beispiel Verkehrszeichen formuliert. Diese wurden hier als gegeben angenommen und sind innerhalb eines umfangreicheren Regelsatzes aufzustellen.

Die formulierten Regeln werden im folgenden Abschnitt exemplarisch auf die zwei betrachteten Szenarien angewendet und das Ergebnis in Bezug auf das inferierte Sollverhalten auf Regelkonformität hin untersucht.

\section{Evaluation der Beispielanwendung}
\label{sec:eval}
Dieser Beitrag betrachtet beispielhaft zwei funktionale Szenarien. Das Szenario, welches eine direkte Einsicht des Ego-Fahrzeugs in den Einlaufbereich des Fußgängerüberwegs ermöglicht, wurde innerhalb der Bottom-up Konzeptualisierung herangezogen. Im Gegensatz dazu wurde das Szenario, welches eine Verdeckung des Einlaufbereichs beinhaltet, nicht bei der Konzeptualisierung betrachtet. Im letzten Schritt der semantischen Normverhaltensanalyse wird das formalisierte Wissen in Form von Konzepten und Regeln in einem Szenarienkatalog geprüft. In diesem Abschnitt wird der formalisierte Regelsatz und daraus resultierende Schlussfolgerungen für das Sollverhalten in den beiden gewählten Szenarien evaluiert.

\begin{figure}[!h]
  \centering
  \captionsetup{width=.8\linewidth}
  \subfloat[][]{\includegraphics[width=0.45\linewidth]{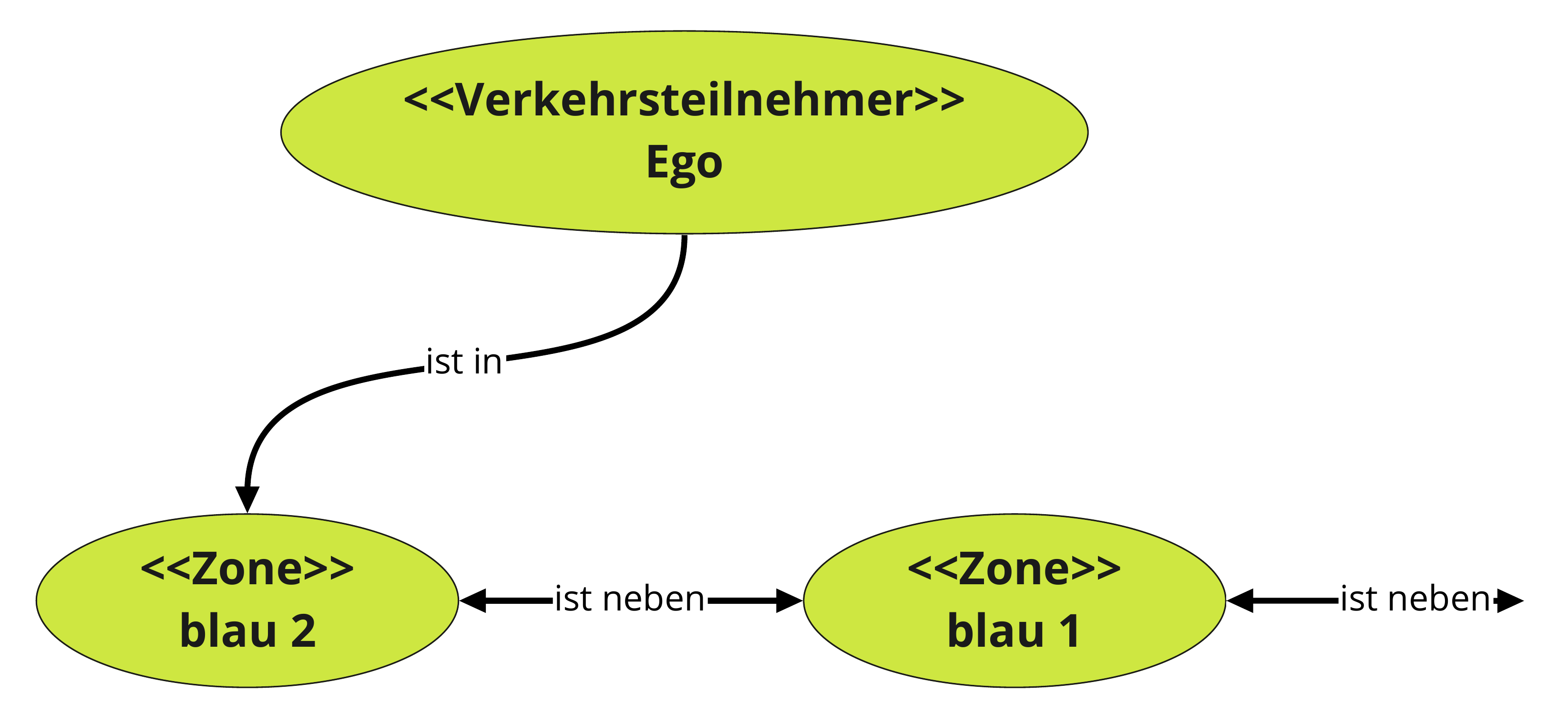}}
  \qquad
  \subfloat[][]{\includegraphics[width=0.45\linewidth]{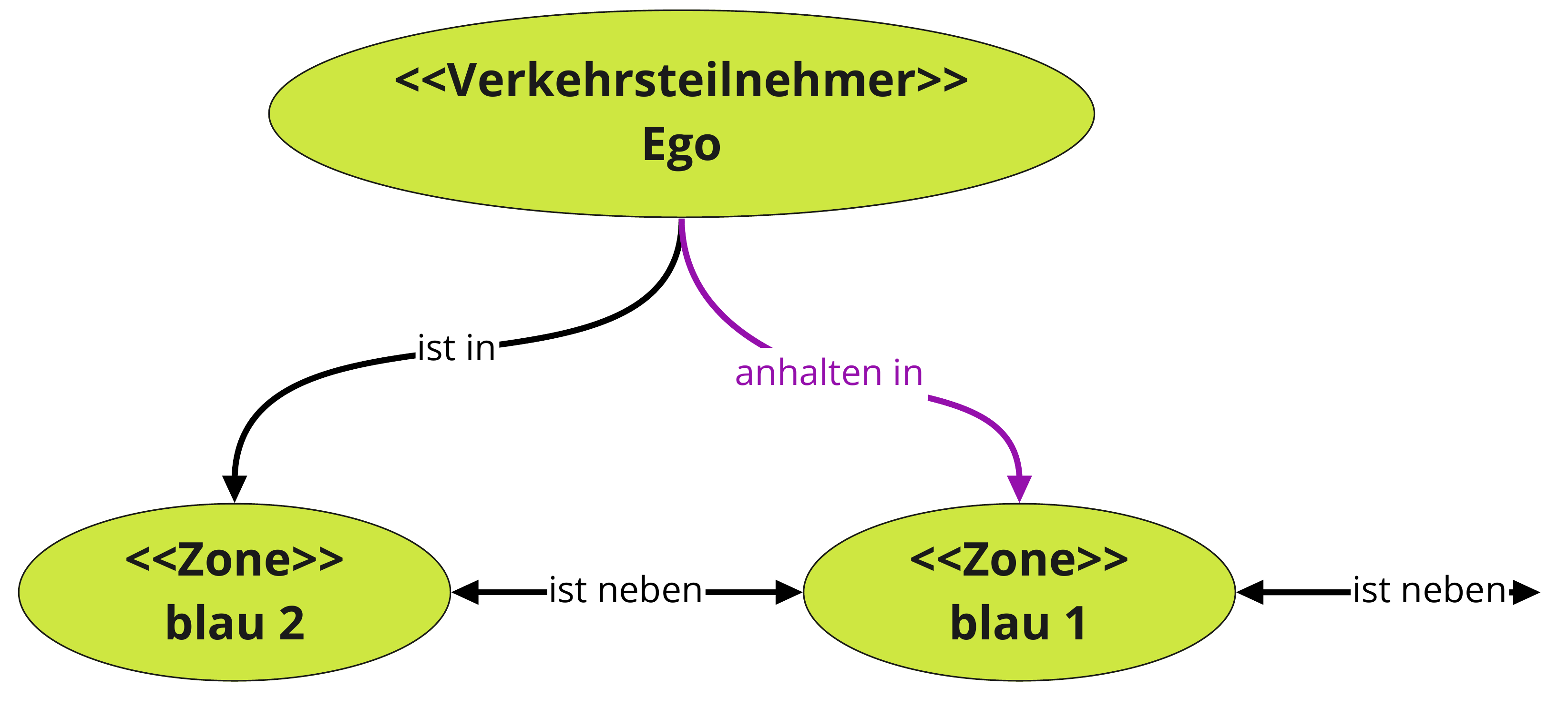}}
  \caption{Die formalisierten Regeln werden auf die in der Szene repräsentierten Entitäten angewendet. Im Anwendungsbeispiel wird auf Basis der Verhaltensregeln die Beziehung \enquote{anhalten\_in} zwischen dem Ego-Fahrzeug und der Zone \enquote{blau 1} inferiert.}\label{fig:inference}
\end{figure}

Innerhalb des Inferenzprozesses werden die in SWRL definierten Regeln auf die Entitäten der ABox der OWL-Ontologie (Ausschnitt in Abbildung \ref{fig:inference}a) angewendet und logische Schlussfolgerungen berechnet. Als Ergebnis wird die Beziehung \enquote{anhalten\_in} zwischen dem Ego-Fahrzeug und der im Beispiel als Zone \enquote{blau 1} bezeichneten Zone inferiert (Abbildung \ref{fig:inference}b). Für die Analyse der Korrektheit der genutzten Regeln soll zukünftig eine Integration des definierten Regelwerks innerhalb eines Phänomen-Signal-Modells~\cite{beck_phanomen-signal-modell_2021, beck_phenomenon-signal_2022} erfolgen.

Als Validierung der Übersetzung der Wissensquellen in die vorgestellte Ontologie und in den dazugehörigen Regelkatalog wird das inferierte Verhalten auf Basis von Expertenwissen und der durchgeführten rechtswissenschaftlichen Analyse überprüft. Hierbei fällt auf, dass das Ego-Fahrzeug bei regelkonformem Verhalten immer vor dem Fußgängerüberweg anhalten würde, wenn ein Fußgänger sich im Einlaufbereich befindet. In diesem verhält\-nis\-mäßig wenig komplexen funktionalen Szenario zeigt sich zum Beispiel bei der Berücksichtigung des erkennbaren Überque\-rungs\-wunsches die Herausforderung der Ableitung und Konkretisierung von Regeln. Unter welchen Bedingungen ein Fußgänger tatsächlich als querungswillig gesehen werden kann, kann beispielsweise Gerichtsurteilen, Feldbeobachtungen oder Prädiktionsmodellen~\cite{bonnin_pedestrian_2014} entnommen werden. Um in diesem Beispiel nicht weitere Wissenquellen einbeziehen zu müssen, wurde die Annahme getroffen, dass ein im Einlaufbereich stehender Fußgänger erkennbar den Fußgängerüberweg queren will. Diese Annahme wird durch die entsprechende Regel explizit dokumentiert.

Die Analyse des zweiten funktionalen Szenarios zeigt eine weitere Herausforderung bei der Skalierbarkeit des Ansatzes innerhalb eines Szenarienkatalogs. Da das zweite Szenario bei der Konzeptualisierung und Regelkonstruktion nicht explizit berücksichtigt wurde, ist das Konzept der Verdeckung (bzw. Zone \enquote{grau 1}) weder in der Ontologie noch in den Regeln berücksichtigt.

Da eine automatisierte Prüfung fehlenden Wissens bisher nicht Teil des Ansatzes ist, bleibt die expertenbasierte Überprüfung aller Szenarien als notwendiger Schritt bestehen. Automatisiert können ausschließlich logische Fehler in der Ontologie und im Regelkatalog festgestellt werden.

\section{Zusammenfassung und Ausblick}
\label{sec:conc}
Diese Arbeit liefert mit der vorgestellten semantischen Normverhaltensanalyse einen Beitrag zur durchgängigen, formalen Verhaltensspezifikation automatisierter Straßenfahrzeuge. Es wurde argumentiert, dass eine explizite Repräsentation des Norm- und Sollverhaltens einen Beitrag zur Absicherung automatisierter Straßenfahrzeuge leisten kann. Auf Basis der Anforderung eines erklärbaren und rückverfolgbaren Verhaltens während des gesamten Lebenszyklus eines automatisierten Fahrzeugs wurde ein Ansatz zur semantischen Normverhaltensanalyse vorgestellt und zu existierenden Ansätzen in der Literatur abgegrenzt.

Die vorgestellte Analyse ist hierbei als Fallbeispiel zu verstehen und soll keine Auslegung der StVO voraussetzen. Im Rahmen der Arbeit wurde ein methodischer Vorschlag erarbeitet, um Expert*innen verschiedener Domänen eine Schnittstelle bereitzustellen, damit Interpretationen von Wissensquellen wie der StVO sinngemäß, formal repräsentiert werden können.

Ein Vorteil des Ansatzes ist, dass die Bewertung des inferierten Sollverhaltens durch Expert*innen explizit in den iterativen Entwicklungsprozess einbezogen werden kann. Da die szenarienbezogene Bewertung als Stärke von Analysen durch den Menschen verstanden wird, kann der Ansatz zur semantischen Normverhaltensanalyse als Unterstützungsmittel zur Herleitung von Verhaltensregeln gesehen werden. Die Stärke der formalen Repräsentation der Verhaltensregeln lässt sich insbesondere unter Berücksichtigung eines umfangreichen Szenarienkatalogs und der damit verbundenen Überprüfung von Widersprüchen im Sollverhalten prognostizieren.

Grenzen des Ansatzes ergeben sich beim Übergang zwischen der nicht-formalen und formalen Beschreibung von Verhalten. Die Interpretation von Verhaltensnormen und die Ableitung von Sollverhalten stellt eine Fehlerquelle dar. Die explizite Dokumentation von Interpretation an diesem Übergang ist wesentlich für die Argumentierbarkeit getroffener Entwurfsentscheidungen. Eine Herausforderung des offenen Kontextes, die der Ansatz nur eingeschränkt adressiert, ist die notwendigerweise getroffene Vereinfachung bei der Erstellung eines Szenarienkatalogs. Die Formulierung von Verhaltensnormen kann zur Argumentierbarkeit von Annahmen in beobachteten kritischen Grenzfällen beitragen, das Treffen der Annahmen selbst aber nicht ausschließen.

Ein wesentlicher, ausstehender technischer Integrationsschritt mit dem im Projekt VVMethoden entwickelten Phänomen-Signal-Modell~\cite{beck_phanomen-signal-modell_2021, beck_phenomenon-signal_2022} steht aus, ist aber in der Konzeption des vorgestellten Ansatzes berücksichtigt. Darüber hinaus bestehen offene Herausforderungen bei der Skalierung des Ansatzes auf weitere Szenarien und Verhaltensregeln zum Beispiel aus der StVO. Zusätzlich müssen weitere Wissenquellen einbezogen werden, um Verhaltensnormen einer Betriebsumgebung valide abdecken zu können. Hierbei ist zukünftig zu untersuchen, inwieweit sich hier beispielhaft angewendete semantische Interpretationen auf andere Rechtsquellen und Rechtskontexte anwenden lassen. Im Zusammenhang der semantischen Kompatibilität und damit verbundenen Interoperabilität der gezeigten und beispielhaft angenommenen Wissensrepräsentation ist eine Harmonisierung mit bestehenden Ontologien ebenfalls Gegenstand zukünftiger Arbeiten.

\section*{Danksagung}
Diese Forschungsarbeiten wurden zum Teil im Rahmen des Projekts „VVMethoden“ durchgeführt. Wir bedanken uns für die finanzielle Unterstützung des Projekts durch das Bundesministerium für Wirtschaft und Klimaschutz (BMWK). Wir bedanken uns zudem bei unseren Kollegen im Projektkonsortium von VVMethoden und insbesondere bei Christian Lalitsch-Schneider (ZF Friedrichshafen AG) und Dr.-Ing.\,Christoph Höhmann (Mercedes-Benz Group AG) für die anregenden Diskussionen, sowie bei Prof.\,Dr.\,rer.\,nat.\,Markus Brandstätter (PROSTEP AG) für viel hilfreiches Feedback. Für die Übernahme des Lektorats danken wir Kim Steinkirchner (PROSTEP AG). Abschließend bedanken wir uns bei Hans Nikolaus Beck (Robert Bosch GmbH), der diese Arbeit fundamental geformt und begleitet hat.

\printbibliography
\end{document}